\documentclass[10pt, sigconf, letterpaper, nonacm]{acmart}

\usepackage{longtable}
\usepackage{array}
\usepackage{booktabs}
\usepackage{listings}
\usepackage{xspace}
\usepackage{xargs}
\usepackage{xstring}
\usepackage{hyperref}
\usepackage[utf8]{inputenc}
\usepackage{supertabular}
\usepackage{multicol}
\usepackage{float}
\usepackage{afterpage}

\usepackage{xcolor}
\definecolor{candypink}{rgb}{0.89, 0.44, 0.48}
\definecolor{beaublue}{rgb}{0.74, 0.83, 0.9}

\usepackage{enumitem}
\setlist[itemize]{leftmargin=0pt}

\newcommand{\abbrEtherscanAddress}[1]{\href{https://etherscan.io/address/#1}{\StrLeft{#1}{6}..\StrRight{#1}{4}}\xspace}

\newcommand{\abbrBscscanAddress}[1]{\href{https://bscscan.com/address/#1}{\StrLeft{#1}{6}..\StrRight{#1}{4}}\xspace}

\title{Do you still need a manual smart contract audit?} 

\usepackage{acro}
\acsetup{single}

\DeclareAcronym{ABI}{
  short = ABI,
  long  = Application Binary Interface,
}

\DeclareAcronym{AMM}{
  short = AMM,
  long  = Automated Market Maker,
}

\DeclareAcronym{API}{
  short = API,
  long  = Application Programming Interface,
}

\DeclareAcronym{BEV}{
  short = BEV,
  long  = Blockchain Extractable Value,
}

\DeclareAcronym{BRF}{
  short = BRF,
  long  = Back-run Flodding,
}

\DeclareAcronym{CeFi}{
  short = CeFi,
  long  = Centralized Finance,
}

\DeclareAcronym{DApp}{
  short = DApp,
  long  = Decentralized Application,
}

\DeclareAcronym{DeFi}{
  short = DeFi,
  long  = Decentralized Finance,
}

\DeclareAcronym{DEX}{
  short = DEX,
  long  = Decentralized Exchange,
}

\DeclareAcronym{DNN}{
  short = DNN,
  long  = Deep Neural Network,
}

\DeclareAcronym{DOS}{
  short = DOS,
  long  = Denial-of-Service,
}

\DeclareAcronym{EV}{
  short = EV,
  long  = Extractable Value,
}

\DeclareAcronym{EVM}{
  short = EVM,
  long  = Ethereum Virtual Machine,
}

\DeclareAcronym{FaaS}{
  short = FaaS,
  long  = Front-running as a Service,
}

\DeclareAcronym{IDS}{
  short = IDS,
  long  = Intrusion Detection System,
}

\DeclareAcronym{ITR}{
  short = ITR,
  long  = Intermediate Trace Representation,
}

\DeclareAcronym{LSTM}{
  short = LSTM,
  long  = Long Short-Term Memory,
}

\DeclareAcronym{MEV}{
  short = MEV,
  long  = Miner Extractable Value,
}

\DeclareAcronym{NLP}{
  short = NLP,
  long  = Natural Language Processing,
}

\DeclareAcronym{P2P}{
  short = P2P,
  long  = peer-to-peer,
}

\DeclareAcronym{PGA}{
  short = PGA,
  long  = Priority Gas Auction,
}

\DeclareAcronym{PoW}{
  short = PoW,
  long  = Proof of Work,
}

\DeclareAcronym{PoS}{
  short = PoS,
  long  = Proof of Stake,
}

\DeclareAcronym{RNN}{
  short = RNN,
  long  = Recurrent Neural Network,
}

\DeclareAcronym{SOTA}{
  short = SOTA,
  long  = State-of-the-Art,
}

\DeclareAcronym{TVL}{
  short = TVL,
  long  = Total Value Locked,
}

\DeclareAcronym{IPS}{
  short = IPS,
  long  = Intrusion Prevention System,
}

\begin{document}

\author{
Isaac David\textsuperscript{1}, 
Liyi Zhou\textsuperscript{2}\textsuperscript{3}, 
Kaihua Qin\textsuperscript{2}\textsuperscript{3}, \\ 
Dawn Song\textsuperscript{3},
Lorenzo Cavallaro\textsuperscript{1},
Arthur Gervais\textsuperscript{1}\textsuperscript{3}}

\affiliation{
	\textsuperscript{1}University College London
    \country{}
	\textsuperscript{2}Imperial College London
    \country{}
	\\
	\textsuperscript{3}UC Berkeley, Center for Responsible Decentralized Intelligence (RDI)
    \country{}
}
\renewcommand{\shortauthors}{David et al.}

\begin{abstract}
We investigate the feasibility of employing large language models (LLMs) for conducting the security audit of smart contracts, a traditionally time-consuming and costly process. Our research focuses on the optimization of prompt engineering for enhanced security analysis, and we evaluate the performance and accuracy of LLMs using a benchmark dataset comprising 52 Decentralized Finance (DeFi) smart contracts that have previously been compromised.

Our findings reveal that, when applied to vulnerable contracts, both GPT-4 and Claude models correctly identify the vulnerability type in 40\% of the cases. However, these models also demonstrate a high false positive rate, necessitating continued involvement from manual auditors. The LLMs tested outperform a random model by 20\% in terms of F1-score.

To ensure the integrity of our study, we conduct mutation testing on five newly developed and ostensibly secure smart contracts, into which we manually insert two and 15 vulnerabilities each. This testing yielded a remarkable best-case 78.7\% true positive rate for the GPT-4-32k model. We tested both, asking the models to perform a binary classification on whether a contract is vulnerable, and a non-binary prompt. We also examined the influence of model temperature variations and context length on the LLM's performance.

Despite the potential for many further enhancements, this work lays the groundwork for a more efficient and economical approach to smart contract security audits.
\end{abstract}

\maketitle

\section{Introduction}\label{sec:introduction}
Decentralized finance has seen a surge in adoption, amplifying the need for robust security measures to guard against the financial consequences of smart contract vulnerabilities. Hundreds of DeFi attacks have led to billions of USD in damages~\cite{zhou2023sok}, underlining the deficiencies of the existing smart contract auditing methodologies in the industry.

This research proposes an innovative approach to improving smart contract auditing by leveraging language models, specifically GPT-4-32k and Claude-v1.3-100k, to identify vulnerabilities within blockchain smart contracts. Despite their inherent limitations, including context truncation and a notable volume of false positives, LLMs exhibit a significant potential in vulnerability detection, achieving a hit rate of 40\% on vulnerable contracts.

Our findings are derived from an exhaustive analysis of 52 vulnerable DeFi smart contracts that have collectively contributed to nearly 1 billion USD in losses. To establish a baseline, we first classified vulnerabilities into distinct types and then engaged the LLMs to interrogate these 38 classes of vulnerabilities. Although manual verification of model outputs and elimination of false positives demands substantial resources, the true value of LLMs lies in their competency to identify genuine vulnerabilities.

To address potential biases originating from the LLM training datasets, we further explore mutation testing. We construct five smart contracts, designed to be secure, and subsequently incorporate two and 15 deliberate vulnerabilities in each. Given that these vulnerabilities are unlikely to be present in the training data of the models, this approach facilitates a more authentic evaluation of the system's robustness and adaptability. In the mutation testing of our five synthetic contracts, we achieve a notable 78.8\% true positive rate.

This paper provides the following key contributions:

\begin{itemize}
    \item To our knowledge, this is the inaugural use of large language models for performing security audits on smart contracts, with a particular focus on the smart contract and DeFi protocol layer. Our research showcases practical prompt engineering methodologies that could enhance and streamline traditional manual smart contract audit processes.
    \item We deliver a quantitative evaluation of the performance and accuracy of two LLMs, GPT-4-32k and Claude-v1.3-100k, against a dataset of 52 DeFi attacks encompassing 38 attack types, related to smart contract vulnerabilities (e.g., reentrancy) and DeFi protocol layer issues (e.g., oracle manipulation attacks). Both models achieve comparable performance, with a 40\% hit rate on vulnerable smart contracts and fewer false positives than a random  baseline model. The LLMs' F1 score is 20\% higher than the random baseline, primarily due to the inflated false positive rate of the latter.
    \item We generate five new supposedly secure contracts, on which we introduce either two or 15 vulnerabilities. We evaluate the vulnerable contracts with a binary classification LLM prompt and a non-binary LLM query. We further study the impact of context length and model temperature on the model performance in smart contract auditing.
    \item We provide two chain-of-thought reasoning case studies employing few-shot prompting. We illustrate the effectiveness of chain-of-thought prompting in identifying recent smart contract vulnerabilities that resulted in losses worth 10M USD, which were unlikely part of the LLM's training dataset.
\end{itemize}

Our research forms an integral stepping stone toward reshaping the smart contract auditing landscape. Given that more than 80\% of the contracts examined fit within GPT-4-32k's context size, the adoption of advanced language models can significantly enhance an auditors' ability to proactively detect and rectify vulnerabilities, thereby strengthening the security fabric of the rapidly evolving DeFi ecosystem.

\section{Background}\label{sec:background}

Since the inception of blockchains with Bitcoin in~$2008$~\cite{bitcoin}, it became apparent that their most well-suited use case is the transfer or trade of financial assets without trusted intermediaries~\cite{wust2018you}. A blockchain is considered permissionless when entities can join and leave the network at any point in time. Users authorize transactions through a public key signature and a subsequent broadcast on the blockchain peer-to-peer network. Due to the openness of the peer-to-peer network, the information about a transaction becomes public, once a transaction is broadcast. For example, blockchain participants can observe which smart contract a pending transaction call triggers along with the corresponding call parameters. Miners, or validators accumulate unconfirmed transactions within blocks to the blockchain. Leader elections can occur via for instance, Proof-of-Work~\cite{bitcoin} or Proof-of-Stake~\cite{saleh2021blockchain,bano2019sok}. In addition to the block reward and transaction fees, Miner Extractable Value is a new miner reward source~\cite{qin2022quantifying}. For a more thorough blockchain background, we refer the reader to several helpful SoKs~\cite{bonneau2015sok,atzei2017survey,bano2017consensus}.

While Bitcoin supports basic smart contracts through a stack-based scripting language, the addition of support for loops and higher-level programming languages (e.g., Solidity) has resulted in widespread adoption. Note that blockchains generally require transaction fees as in to prevent Denial of Service attacks. Smart contracts are therefore only quasi Turing-complete because their execution can suddenly interrupt if the transaction fees exceed a predefined amount. Notably, blockchains do not store the human-readable source code, nor the application interface to interact with a smart contract (i.e., the Application Binary Interface). Instead, blockchains store the compiled bytecode on-chain.


DeFi refers to an ecosystem of financial products and services built on top of permissionless blockchains. DeFi has experienced a surge in popularity, with a peak Total Value Locked (TVL) \href{https://defillama.com/}{reaching $250B$~USD in December~$2021$}. Despite incorporating basic functions inspired by traditional finance (e.g., lending~\cite{qin2021empirical}, trading, derivatives), DeFi also introduces novel designs (e.g., flash loans~\cite{qin2021attacking}, composable trading~\cite{zhou2021just}) enabled by a blockchain's atomicity property and brought automated market makers~\cite{zhou2021high} to popularity.
Understanding the semantics of transactions that trigger DeFi designs presents a particular challenge because DeFi transactions typically are intertwined with multiple financial decentralized applications.

\section{Natural Language Processing}
Natural language models are machine learning models that are designed to process and generate human-like text. They are used in a wide range of applications, including language translation, text summarization, language generation, and text classification.

\paragraph*{NLP and Bytecode}
There have been several approaches to applying natural language models to code, assembly, and bytecode. One approach is to treat the code or assembly as natural language text and input it into a natural language model. This can be useful for tasks such as code summarization, code generation, or code translation. Another approach is to first convert the code or assembly into a structured representation, such as an abstract syntax tree (AST), and then input the AST into a natural language model. 

Bytecode, which is a low-level representation of code that is typically executed by a virtual machine, can also be processed using natural language models. One approach is to convert the bytecode into a higher-level representation, such as assembly code, and then input it into a natural language model. Another approach is to treat the bytecode as a sequence of tokens and input it into a sequence-to-sequence natural language model, which can be useful for tasks such as bytecode translation or bytecode summarization.

\paragraph*{Generative Pre-trained Transformer} GPT is a language model developed by OpenAI that uses deep learning techniques to generate human-like text. GPT-1, released in 2018, is the original version of the model, trained on a dataset of 8 million web pages. GPT-1 can generate coherent paragraphs of text on a variety of topics. GPT-2, from 2019, is an updated version of the model~\cite{radford2019language}. It was trained on a much larger dataset of internet text, including books, articles, and websites, and can generate more sophisticated and human-like text. GPT-4, from 2023, is the latest version of the model. It is trained on an even larger dataset than GPT-3~\cite{brown2020language}, an autoregressive language model with 175 billion parameters. GPT-4 provides strong task-agnostic few-shot performance via text interaction with the model, beating in some cases state-of-the-art fine-tuning approaches for specific tasks.

\paragraph*{Claude}
Claude is an AI assistant developed by Anthropic, an American artificial intelligence startup consisting of former OpenAI members. It focuses on conversational and text processing tasks, such as summarization, search, Q\&A, and coding. Claude is accessible through a chat interface and API, offering users the ability to customize its personality, tone, and behavior. As a next-generation AI model, it aims to provide improved user interaction and reliable output.

\section{Data Collection}
Applying LLMs to smart contract security first requires a suitable attack dataset, the respective vulnerabilities and the corresponding smart contract source code. Based on the DeFi Attack SoK~\cite{zhou2023sok}, we identify 52 attacked DeFi protocols (cf.\ Table~\ref{tab:ground-truth}), for which we can also find the associated vulnerable source code. The total damage in USD caused by these attacks is estimated to reach 956M USD. For each attack, we then categorize the respective vulnerabilities among 38 distinct vulnerability types (cf.\ Table~\ref{tab:attack-types} and~\ref{tab:attack-types2}). For each vulnerable smart contract, we collect the source code from the \href{https://docs.etherscan.io/}{Etherscan API}. Etherscan entertains a set of verified smart contracts, as such, under the assumption that Etherscan can be a reliable third party, we can be confident that the source code matches the attacked smart contracts.

\paragraph*{LLM Context Limits}
Large language models have context limitation, i.e.\ the prompt provided to an LLM is limited by the number of tokens. In June 2023, GPT-4 has a default token limit of 8000 tokens, while Claude offers up to 100k tokens within its context. For this study, we are fortunate to have access to both, Claude's v1.3-100k model and GPT-4's 32k model with a limit of 32000 tokens. Such context space is important due to the following reasons:

\begin{enumerate}
    \item Ideally, the entire source code that should be analyzed can be provided as is to the LLM as context. If the source code does not fit, we must necessarily split or truncate the source code and provide its content in separate queries, which is suboptimal.
    \item The audit context, i.e., the context we provide to the LLM to find a particular vulnerability, shall be sufficiently explicit for the LLM to perform well. As such, in addition to the source code's token length, the audit context may occupy token space.
\end{enumerate}

\begin{figure}[htb]
\begin{center}
\includegraphics[width=\columnwidth]{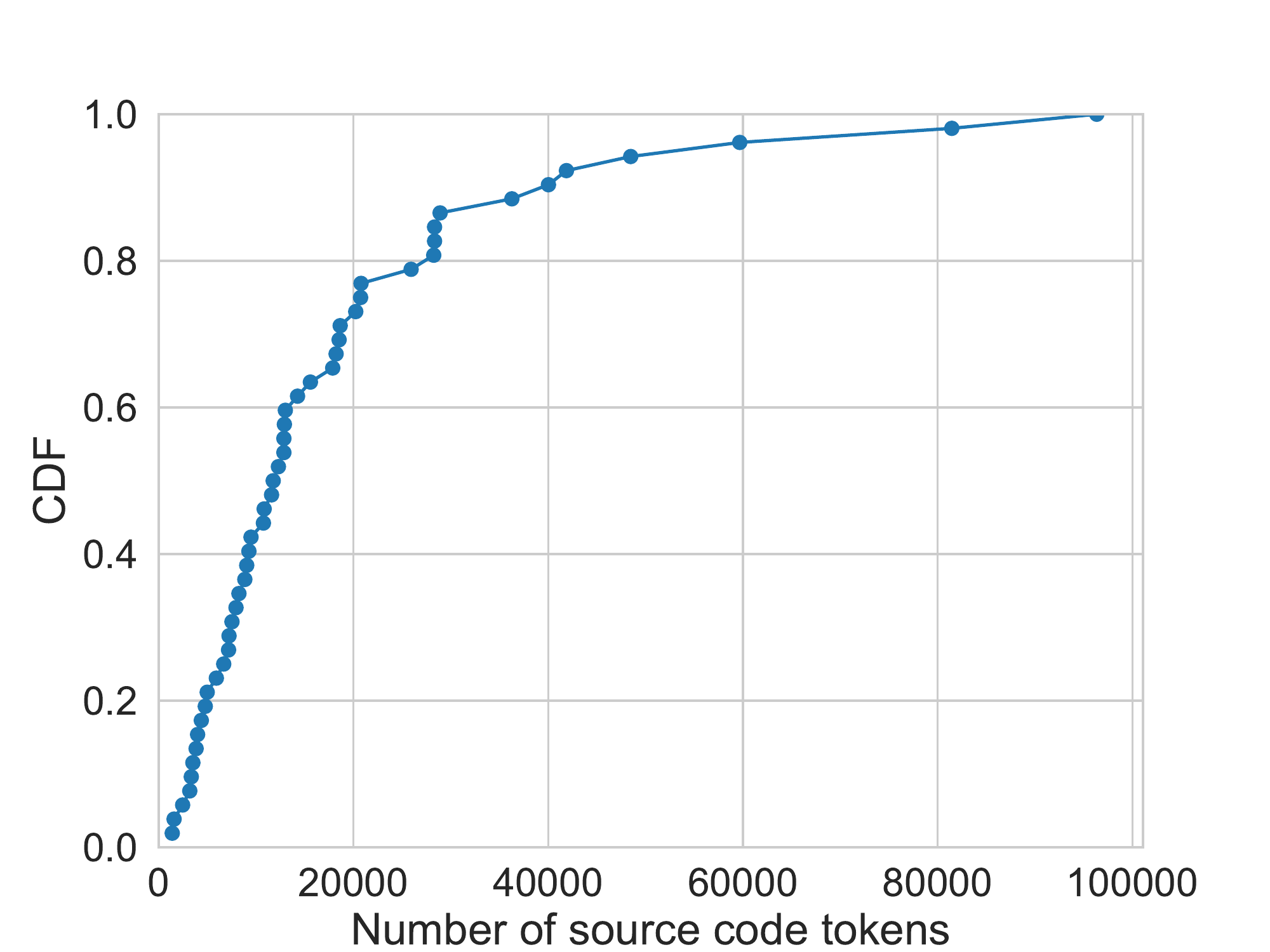}
\caption{Distribution of the token lengths of the 52 attacked smart contracts. Most of the vulnerable smart contracts fit into the context length of GPT-4-32k.}
\label{fig:code-token-length-distribution}
\end{center}
\end{figure}

To assess whether the LLM's at hand are sufficiently adept at covering the vulnerable contract's context, we plot in Figure~\ref{fig:code-token-length-distribution} their respective length. We find that nearly 90\% of the analyzed contracts are less than 30k tokens in length, hence fitting into both, GPT-4-32k and Claude-v1.3-100k. Only 7 of the analyzed contracts exceed GPT-4-32k's token limit, while they still fit into Claude. Note that for this paper, we limit the source code of GPT-4 to 31k tokens, and the source code passed to Claude to 99k tokens. As such, we leave space for 1000 tokens for the prompt engineering.

\begin{table*}
\centering
\caption{52 victim smart contract addresses and attack type ground truth extracted from the DeFi Attack SoK~\cite{zhou2023sok}.}
\label{tab:ground-truth}
\begin{multicols}{2}
    \begin{tabular}{>{\raggedright}p{2cm} p{6cm}}
        \toprule
        \textbf{Address} & \textbf{Attack Type Ground Truth}\\
        \midrule
        \abbrEtherscanAddress{0xe952bda8c06481506e4731C4f54CeD2d4ab81659} & Reentrancy, Token standard incompatibility \\ \addlinespace
\abbrEtherscanAddress{0x55dBb68f69337FCABe261f296B50B4723D333830} & On-chain oracle manipulation, Absence of code logic or sanity check \\ \addlinespace
\abbrEtherscanAddress{0x4e3FBD56CD56c3e72c1403e103b45Db9da5B9D2B} & Governance attack \\ \addlinespace
\abbrEtherscanAddress{0x833e440332cAA07597a5116FBB6163f0E15F743D} & On-chain oracle manipulation \\ \addlinespace
\abbrEtherscanAddress{0x74BCb8b78996F49F46497be185174B2a89191fD6} & Absence of code logic or sanity check \\ \addlinespace
\abbrEtherscanAddress{0x0624Eb9691D99178d0d2Bd76c72f1dBB4DB05286} & On-chain oracle manipulation \\ \addlinespace
\abbrEtherscanAddress{0x7B3B69eAb43C1AA677DF04b4B65F0D169Fcdc6Ca} & Visibility errors, including unrestricted action \\ \addlinespace
\abbrEtherscanAddress{0x3212b29E33587A00FB1C83346f5dBFA69A458923} & Token standard incompatibility, Reentrancy \\ \addlinespace
\abbrEtherscanAddress{0x5bD628141c62a901E0a83E630ce5FaFa95bBdeE4} & Unfair slippage protection, Absence of code logic or sanity check \\ \addlinespace
\abbrEtherscanAddress{0x32e574b0400CdB93DC33A34a30986737186Bde43} & Flash liquidity borrow, purchase, mint or deposit \\ \addlinespace
\abbrEtherscanAddress{0x39b1dF026010b5aEA781f90542EE19E900F2Db15} & On-chain oracle manipulation, Frontrunning \\ \addlinespace
\abbrEtherscanAddress{0x17e8Ca1b4798B97602895f63206afCd1Fc90Ca5f} & Deployment mistake, Other unsafe DeFi protocol dependency \\ \addlinespace
\abbrEtherscanAddress{0x923cAb01E6a4639664aa64B76396Eec0ea7d3a5f} & Visibility errors, including unrestricted action \\ \addlinespace
\abbrEtherscanAddress{0xC1E088fC1323b20BCBee9bd1B9fC9546db5624C5} & Governance attack \\ \addlinespace
\abbrEtherscanAddress{0x88093840AaD42d2621e1a452BF5d7076fF804D61} & Token standard incompatibility \\ \addlinespace
\abbrEtherscanAddress{0x3Ec4A6Cfe803ee84009Ce6E1ecF419C9Cb1E8aF0} & On-chain oracle manipulation \\ \addlinespace
\abbrEtherscanAddress{0x33bf0Bb8E1405dc440ECcB97Ffd92fef438c8a27} & Other fake contracts, Deployment mistake,Arithmetic mistakes \\ \addlinespace
\abbrEtherscanAddress{0x25a5feB5aC6533fE3C4E8E8e2a55f9E1f1F8E5f0} & Absence of code logic or sanity check \\ \addlinespace
\abbrEtherscanAddress{0x951D51bAeFb72319d9FBE941E1615938d89ABfe2} & Absence of code logic or sanity check \\ \addlinespace
\abbrEtherscanAddress{0xE0B94a7BB45dD905c79bB1992C9879f40F1CAeD5} & Other coding mistakes \\ \addlinespace
\abbrEtherscanAddress{0x67B66C99D3Eb37Fa76Aa3Ed1ff33E8e39F0b9c7A} & Other unsafe DeFi protocol dependency, Absence of code logic or sanity check \\ \addlinespace
\abbrEtherscanAddress{0x9Daec8D56CDCBDE72abe65F4a5daF8cc0A5bF2f9} & Other unsafe DeFi protocol dependency, Inconsistent access control \\ \addlinespace
\abbrEtherscanAddress{0x0e511Aa1a137AaD267dfe3a6bFCa0b856C1a3682} & Other unsafe DeFi protocol dependency \\ \addlinespace
\abbrEtherscanAddress{0x75572098dc462F976127f59F8c97dFa291f81d8b} & Visibility errors, including unrestricted action \\ \addlinespace
\abbrEtherscanAddress{0x6847259b2B3A4c17e7c43C54409810aF48bA5210} & Absence of code logic or sanity check, Other unsafe DeFi protocol dependency \\ \addlinespace
\abbrEtherscanAddress{0x6684977bBED67e101BB80Fc07fCcfba655c0a64F} & Other unsafe DeFi protocol dependency \\ \addlinespace

    \end{tabular}
    
\columnbreak

    \begin{tabular}{>{\raggedright}p{2cm} p{6cm}}
        \toprule
        \textbf{Address} & \textbf{Attack Type Ground Truth}\\
        \midrule
        \abbrEtherscanAddress{0x35C674C288577Df3e9b5dafEF945795b741c7810} & Flash liquidity borrow, purchase, mint or deposit, Absence of code logic or sanity check \\ \addlinespace
\abbrEtherscanAddress{0x66e7d7839333f502df355f5bd87aEA24BAC2eE63} & Absence of code logic or sanity check \\ \addlinespace
\abbrEtherscanAddress{0x55dBb68f69337FCABe261f296B50B4723D333830} & On-chain oracle manipulation, Absence of code logic or sanity check \\ \addlinespace
\abbrEtherscanAddress{0xDdD7df28B1Fb668B77860B473aF819b03DB61101} & On-chain oracle manipulation \\ \addlinespace
\abbrEtherscanAddress{0xC9f27A50f82571C1C8423A42970613b8dBDA14ef} & Reentrancy, Visibility errors, including unrestricted action, Delegatecall injection \\ \addlinespace
\abbrEtherscanAddress{0x0eEe3E3828A45f7601D5F54bF49bB01d1A9dF5ea} & Token standard incompatibility, Reentrancy \\ \addlinespace
\abbrEtherscanAddress{0x88Cc4aA0dd6Cf126b00C012dDa9f6F4fd9388b17} & Absence of code logic or sanity check \\ \addlinespace
\abbrEtherscanAddress{0xae465FD39B519602eE28F062037F7B9c41FDc8cF} & On-chain oracle manipulation \\ \addlinespace
\abbrEtherscanAddress{0x818E6FECD516Ecc3849DAf6845e3EC868087B755} & On-chain oracle manipulation \\ \addlinespace
\abbrEtherscanAddress{0xAcbdB82f07B2653137d3A08A22637121422ae747} & Flash liquidity borrow, purchase, mint or deposit \\ \addlinespace
\abbrEtherscanAddress{0x6b7a87899490EcE95443e979cA9485CBE7E71522} & Unsafe call to phantom function, Absence of code logic or sanity check \\ \addlinespace
\abbrEtherscanAddress{0xA23179BE88887804f319C047E88FDd4dD4867eF5} & Direct call to untrusted contract, Insider trade or other activities \\ \addlinespace
\abbrEtherscanAddress{0x929cb86046E421abF7e1e02dE7836742654D49d6} & Delegatecall injection, Absence of code logic or sanity check \\ \addlinespace
\abbrEtherscanAddress{0x3157439C84260541003001129c42FB6aBa57E758} & Governance attack \\ \addlinespace
\abbrEtherscanAddress{0x0e6ffd4dAecA13A8158146516f847D2F44AD4A30} & Locked or frozen tokens \\ \addlinespace
\abbrEtherscanAddress{0x606e9758a39d2d7fA7e70BC68E6E7D9b02948962} & Absence of code logic or sanity check \\ \addlinespace
\abbrEtherscanAddress{0x1ceC0e358f882733c5ecC028d8A0c24Baee02004} & Reentrancy \\ \addlinespace
\abbrEtherscanAddress{0x876B9EbD725d1Fa0b879FCEe12560a6453b51Dc8} & Visibility errors, including unrestricted action \\ \addlinespace
\abbrEtherscanAddress{0x328d15F6B5Eba1C30CDe1A5F1f5A9E35b07f5424} & Reentrancy, Flash liquidity borrow, purchase, mint or deposit \\ \addlinespace
\abbrEtherscanAddress{0x77f973FCaF871459aa58cd81881Ce453759281bC} & Other unfair or unsafe DeFi protocol interaction, Absence of code logic or sanity check \\ \addlinespace
\abbrEtherscanAddress{0xdE744d544A9d768e96C21B5F087Fc54b776E9b25} & Absence of code logic or sanity check \\ \addlinespace
\abbrEtherscanAddress{0xdAC17F958D2ee523a2206206994597C13D831ec7} & Absence of code logic or sanity check \\ \addlinespace
\abbrEtherscanAddress{0xe7f445B93eB9CDaBfe76541Cc43Ff8dE930A58E6} & Absence of code logic or sanity check \\ \addlinespace
\abbrEtherscanAddress{0x0EFB384d843A191c02F5C4470D0f9EC0122a1c0b} & Absence of code logic or sanity check, Fake tokens \\ \addlinespace
\abbrEtherscanAddress{0xE11fc0B43ab98Eb91e9836129d1ee7c3Bc95df50} & Unfair slippage protection \\ \addlinespace
\abbrEtherscanAddress{0xACd43E627e64355f1861cEC6d3a6688B31a6F952} & Flash liquidity borrow, purchase, mint or deposit, Unfair liquidity providing \\ \addlinespace
    \end{tabular}
\end{multicols}
\end{table*}

\subsection{Covered Attack Types}
In this work, we cover 38 attacks outlined in Table~\ref{tab:attack-types} and~\ref{tab:attack-types2} in the Appendix, along with their respective descriptions that we pass on to the language model context. 

\section{Evaluating GPT-4 and Claude on 52 DeFi Attacks}\label{sec:52-evaluation}
We evaluate our prompt engineering approach by invoking the GPT-4-32k and Claude-v1.3-100k API on each of the 52 DeFi attacks. For each attack, we perform 38 independent runs to account for all the covered attack types in Table~\ref{tab:attack-types}, resulting in a total of $52\times 38 \times 2 = 3'952$ API queries. To avoid manual investigations, for each vulnerability type, we ask the model to perform a binary classification by returning either ``YES'' or ``NO'', respectively, indicating whether this type of vulnerability is present in the provided smart contract. All LLM invocations are run as a single-shot example, i.e.\ we reinitialize the model for each request. Our generic prompt is as follows (cf.\ Listing~\ref{lst:generic-binary-prompt}).

\begin{lstlisting}[breaklines, basicstyle=\footnotesize, caption={Binary classification LLM prompt to identify smart contract vulnerabilities.}, label={lst:generic-binary-prompt}]
    You are an AI smart contract auditor that excels at finding vulnerabilities in blockchain smart contracts. Review the following smart contract code in detail and very thoroughly. Think step by step, carefully. Is the following smart contract vulnerable to '{vulnerability_type}' attacks? Reply with YES or NO only. Do not be verbose. Think carefully but only answer with YES or NO! To help you, find here a definition of a '{vulnerability_type}' attack: {vulnerability_description}\n\n
    Source code: {source_code}
\end{lstlisting}

After each API query, we artificially pause our code for 13 seconds, not to hit API rate limits imposed by e.g., OpenAI. 

Regarding costs, the Claude API was for free for us at the time of writing. For GPT-4-32k, we had generously received free credits from OpenAI and used about 2000 USD for the evaluation on the 52 DeFi attacks.

\subsection{Overall Performance}
When merging Claude and GPT's performance, we find that 58 out of the 146 vulnerability types on 52 DeFi attacks are identified, corresponding to a hit rate of 40\%. However, this result costs a total of, 1318 false positives. In practice, these false positives require costly manual labor to identify whether the vulnerabilities are true positives (cf.\ Table~\ref{tab:tp-fp-etc}).

\begin{table}[H]
\centering
\begin{tabular}{|l|c|c|c|c|}
\hline
\textbf{Model} & \textbf{TP} & \textbf{TN} & \textbf{FP} & \textbf{FN} \\
\hline
claude-v1.3-100k & 26 & 1290 & 578 & 47 \\
gpt-4-32k & 32 & 1128 & 740 & 41 \\
random & 36.54 & 960.62 & 962.42 & 39.21 \\
\hline
\end{tabular}
\caption{False positives/negative and true positives/negatives of GPT-4-32k (default temperature = 0.7), Claude-v1.3-100k (default temperature = 1) and a random model on 52 DeFi attacks, causing total financial damage of 956M USD. Claude and GPT-4 outperform the random baseline in particular on the number of false positives. The random baseline has between 282 and 387 more false positives than the LLMs.}
\label{tab:tp-fp-etc}
\end{table}

We find that although GPT-4 has a shorter context, which requires us to truncate the smart contract source code, GPT-4-32k has a better overall F1-score of 0.077 compared to 0.076 for Claude-v1.3-100k. While the number of false positives is a significant cost in practice, since auditors will have to manually sift through their merit, we find that 4.5\% of the positives that the LLMs finds, are in fact, true positives. We may also want to highlight that simply identifying that a vulnerability class exist isn't sufficient to finalize an audit. An audit ideally requires a precise description of the issue at hand, coupled with a Proof of Concept implementation of the exploit targeting the vulnerability.

\subsection{Considerations}
We believe that this work is just the beginning of a revolution in smart contract auditing. Our work certainly has a set of limitations that we hope will be addressed by future work.

\begin{description}
    \item[Training Data] The LLMs we evaluate within this work are black boxes and the precise training datasets are unknown. As such, we cannot rule out that the LLMs have been either partially or fully trained on some of the attacks we query the models on. To reduce any biases resulting out of this limitation, we perform mutation testing on newly written contracts in Section~\ref{sec:mutation}. 
    \item[Binary Classification] We notice that both, GPT-4 and Claude do not always obey to perform binary classification. Under some instances, the model may, for example, respond with statements such as ``the smart contract is vulnerable to X'', while occluding the keyword YES or NO from the answer. We therefore perform manual verifications on the model text outputs to make sure that our metrics remain coherent.
    \item[False Positives] Not that we do not know the ultimate ground truth of the 52 vulnerable DeFi contracts. These contracts could, theoretically, contain more vulnerabilities than were reported, and hence, the false positive rate of the LLMs could be lower.
    \item[Truncation] We find that seven of the smart contracts source codes exceed the context limit from GPT-4-32k. As such, we perform a naive truncation of these longer source codes to match 31k tokens. A more sophisticated approach may be to compress the source code, or to perform multiple chained queries, which we leave for future work.
    \item[Longer Context] We have noticed qualitatively that the longer the provided context, the less likely Claude will perform binary classification as requested at the beginning of the prompt. We are not aware of the internal operations from Claude to enable a 100k context, but it would be interesting for future research to quantify to what degree a longer context may deteriorate the requests to the model.
\end{description}

\section{Mutation Testing}\label{sec:mutation}
The 52 DeFi attacks used for our evaluation in Section~\ref{sec:52-evaluation} are derived from publicly reported and well-documented incidents. Therefore, it is plausible that GPT-4 and Claude might have been trained on datasets including descriptions of these attacks. To account for this potential bias and to further scrutinize the performance of the models, we employ an additional testing methodology known as mutation testing.

In this method, we generate five unique smart contracts, each designed to be (hopefully) secure. We then introduce a series of vulnerabilities into these contracts and query both GPT-4 and Claude using the same prompt as provided in Section~\ref{sec:52-evaluation}. The initial code for the first contract, intended for mutation testing, is available in Listing~\ref{lst:secure-mutation}. We operate under the assumption that this contract is secure and devoid of any vulnerabilities. Subsequently, we introduce either 2 or 15 distinct vulnerabilities, into the secure contracts. We direct the interested reader to the Appendix for the full code in Listing~\ref{lst:full-mutation-code},~\ref{lst:full-mutation-code-2},~\ref{lst:full-mutation-code-3},~\ref{lst:full-mutation-code-4} and~\ref{lst:full-mutation-code-5}. The comments in the code clarify which vulnerabilities are introduced in which lines of code.

Following the insertion of vulnerabilities, we query both Claude and GPT-4 using the binary prompts from Section~\ref{sec:52-evaluation} for each vulnerability. Specifically, we ask the models to perform a binary classification to determine the presence or absence of the respective vulnerabilities. We therefore perform 760 API calls for the mutation testing: 2 (one for each model) x 38 vulnerability types x 5 contracts x 2 (either 2 or 15 vulnerabilities).

In addition to the binary queries, we also perform an independent experiment with a non-binary system message with the following prompt (cf.\ Listing~\ref{lst:mutation-non-binary-prompt}). The API costs for the mutation testing amounts to about 400 USD.

\begin{lstlisting}[caption={Non-binary prompt for mutation testing.}, label={lst:mutation-non-binary-prompt}]
System Instruction: You are an AI smart contract auditor that excels at finding vulnerabilities in blockchain smart contracts. Review the following smart contract code in detail and very thoroughly.

{{contract source code}}
\end{lstlisting}

\begin{lstlisting}[caption={Supposedly secure contract example for mutation testing, on which we introduce 2 and 15 vulnerabilities (see Listing~\ref{lst:full-mutation-code}).}, label={lst:secure-mutation}]
import "@openzeppelin/contracts-ethereum-package/contracts/token/ERC20/ERC20.sol";
import "@openzeppelin/contracts-ethereum-package/contracts/token/ERC20/IERC20.sol";
import "@openzeppelin/contracts-ethereum-package/contracts/token/ERC20/extensions/IERC20Metadata.sol";
import "@openzeppelin/contracts-ethereum-package/contracts/utils/Context.sol";
import "@openzeppelin/contracts-ethereum-package/contracts/access/Ownable.sol";

contract AirdropFaucet is Ownable {
    uint256 public dripAmount;
    mapping(address => bool) public previousRequestors;
    address public tokenAddress;

    event TokensRequested(address requestor);

    constructor(address _tokenAddress) {
        tokenAddress = _tokenAddress;
        uint8 decimals = ERC20(tokenAddress).decimals();
        dripAmount = 1000 * 10 ** decimals;
    }

    function requestTokens(address requestor) external {
        require(
            previousRequestors[requestor] == false,
            "Address has already requested tokens"
        );
        ERC20 faucetToken = ERC20(tokenAddress);
        require(
            faucetToken.balanceOf(address(this)) > dripAmount,
            "Insufficient funds in faucet"
        );

        // Transfer token to requestor
        faucetToken.transfer(requestor, dripAmount);
        previousRequestors[requestor] = true;

        emit TokensRequested(requestor);
    }

    function withdrawTokens() external onlyOwner {
        ERC20 faucetToken = ERC20(tokenAddress);
        uint256 amount = faucetToken.balanceOf(address(this));
        faucetToken.transfer(msg.sender, amount);
    }

    function setDripAmount(uint256 _dripAmount) external onlyOwner {
        dripAmount = _dripAmount;
    }

    function setTokenAddress(address _tokenAddress) external onlyOwner{
        tokenAddress = _tokenAddress;
    }
}
\end{lstlisting}

\subsection{Mutation Testing Results}
The results for GPT-4 and Claude on the binary and non-binary classification prompts follow.

\paragraph*{Binary Prompt}
In Table~\ref{tab:binary-2-vuln}, where the contracts contain each 2 vulnerabilities, Claude shows a better performance in terms of true positives  when compared to GPT-4 across all contracts. However, both GPT-4 and Claude deliver results with lower false-positive rates than the Random method, though GPT-4 had a higher number of false positives  than Claude in most contract scenarios.

\begin{table}[htb]
\centering
\caption{Binary classification result for contracts containing 2 vulnerabilities.}
\begin{tabular}{@{}ccccccc@{}}
\toprule
\textbf{Contract Name} & \multicolumn{2}{c}{\textbf{GPT-4}} & \multicolumn{2}{c}{\textbf{Random}} & \multicolumn{2}{c}{\textbf{Claude}} \\
\cmidrule(lr){2-3} \cmidrule(lr){4-5} \cmidrule(lr){6-7}
& TP & FP & TP & FP & TP & FP \\ \midrule
LPManager      & 1 & 3 & 1 & 18 & 1 & 7 \\
PriceUpdater   & 0 & 9 & 2 & 22 & 1 & 10 \\
FaucetToken    & 0 & 7 & 2 & 20 & 0 & 3 \\
DAOVoting      & 1 & 13 & 0 & 22 & 2 & 6 \\
AirdropFaucet  & 0 & 6 & 0 & 21 & 0 & 10 \\ \midrule
\textbf{Ground Truth}   & \textbf{2} & \textbf{0} & \textbf{2} & \textbf{0} & \textbf{2} & \textbf{0} \\ \bottomrule
\label{tab:binary-2-vuln}
\end{tabular}
\end{table}

Table~\ref{tab:binary-15-vuln} shows the results of binary classification for contracts containing fifteen vulnerabilities. We can deduce that GPT-4 exhibits a better performance in terms of successfully identifying true positives  across different contracts than Claude, once again Claude demonstrated lower False Positive rates in most scenarios. The Random method results in a significantly higher number of false positives  when compared to both GPT-4 and Claude.

\begin{table}[htb]
\centering
\caption{Binary classification result for contracts containing 15 vulnerabilities.}
\begin{tabular}{@{}cccccccc@{}}
\toprule
\textbf{Contract Name} & \multicolumn{2}{c}{\textbf{GPT-4}} & \multicolumn{2}{c}{\textbf{Random}} & \multicolumn{2}{c}{\textbf{Claude}} \\
\cmidrule(lr){2-3} \cmidrule(lr){4-5} \cmidrule(lr){6-7}
& TP & FP & TP & FP & TP & FP \\ \midrule
LPManager      & 12 & 8 & 7 & 16 & 5 & 4 \\
PriceUpdater   & 12 & 12 & 6 & 12 & 6 & 5 \\
FaucetToken    & 12 & 6 & 10 & 16 & 4 & 5 \\
DAOVoting      & 11 & 7 & 5 & 11 & 4 & 5 \\
AirdropFaucet  & 12 & 4 & 6 & 10 & 7 & 6 \\ \midrule
\textbf{Ground Truth} & \textbf{15} & \textbf{0} & \textbf{15} & \textbf{0} & \textbf{15} & \textbf{0} \\ \bottomrule
\label{tab:binary-15-vuln}
\end{tabular}
\end{table}

From these numbers, it becomes apparent that the smaller the ratio of the number of ground truth vulnerabilities over the number of tested vulnerabilities is, the worse the models perform. This insight is interesting because while we increase the number of tested vulnerabilities, we can expect the number of false positives to also increase. As such, to minimize the human labor cost on verifying positives from the models, it is important to keep the number of tested vulnerabilities to a minimum. However, there's an inherent tension that extending the tested vulnerabilities may help to find more true positives.



\paragraph*{Non-Binary Prompt}
Table~\ref{tab:comparison} compares GPT and Claude's true positive performance on a non-binary test of 5 smart contracts containing 2 and 15 vulnerabilities, respectively. Note that in this experiment, we are not reporting false positives as we focus on examining the open-ended responses of the language models. For contracts with 2 vulnerabilities, GPT-4  identifies the ground truth count of vulnerabilities across all contracts except LPManager where no vulnerabilities are identified. Claude can identify the ground truth number of vulnerabilities in most contracts except FaucetToken, with no vulnerability detected. Both LLMs detect 6 out of 10 vulnerabilities (total vulnerabilities across all the contracts). For contracts with 15 vulnerabilities, GPT-4 performs slightly better and can identify 46 out of the possible 75 vulnerabilities (61.3\%)  in all 5 contracts, while Claude identifies 44 out of these 75 possible vulnerabilities (58.6\%)  correctly.

\begin{table}[htb]
\centering
\caption{Comparison of GPT and Claude's true positives  on a non-binary test of 5 smart contracts with 2 and 15 vulnerabilities, respectively.}
\label{tab:comparison}
\begin{tabular}{lcc}
\toprule
Smart Contract         & GPT TP & Claude TP \\
\midrule
2 vulnerabilities/contract & & \\ 
\midrule
\textbf{Ground Truth}  & \textbf{2}          & \textbf{2}            \\
LPManager      & 0               & 1                  \\
DAOVoting      & 2               & 2                  \\
PriceUpdater   & 2               & 2                  \\
AirdropFaucet  & 1               & 1                  \\
FaucetToken    & 1               & 0                  \\
\midrule
15 vulnerabilities/contract & & \\ 
\midrule
\textbf{Ground Truth}  & \textbf{15}          & \textbf{15}            \\
LPManager      & 7               & 6                  \\
DAOVoting      & 10               & 11                  \\
PriceUpdater   & 6               & 9                  \\
AirdropFaucet  & 11               & 10                  \\
FaucetToken    & 12               & 8                  \\

\bottomrule
\end{tabular}
\end{table}

\paragraph*{Binary or Non-Binary}
We observe that the non-binary prompt tends to deliver better results in terms of true positives, compared to the binary prompt for both GPT-4 and Claude for the contracts with 2 vulnerabilities. For contracts with 15 vulnerabilities, GPT-4 non-binary prompt achieves 61.3\% true positives, compared to the binary prompt results of 78.7\% true positives. However, improved results are seen for Claude in the non-binary setting, achieving a higher true positive rate at 58.6\% in comparison to 34.7\% in the binary setting. Our results suggest that utilizing a non-binary open-ended approach to describe the vulnerabilities would lead to extracting more informative responses from the language models. We believe this is because the non-binary prompt allows models to explore and explain different vulnerabilities in detail, avoiding the limiting nature of the binary choice, and thus potentially increasing the chances of identifying vulnerabilities. The downside of using the non-binary prompt is the need for manual verification of identified vulnerabilities for larger scale studies, which may increase the time and effort required by a human auditor.

\subsection{Performance under Varying Parameters}
In addition to the previous mutation testing, we specifically focus on the AirdropFaucet contract (cf.\ Listing~\ref{lst:full-mutation-code}) and explore the effects of changing model temperature and context length on the performance of GPT-4 and Claude in identifying vulnerabilities within this contract. These parameters can play a crucial role in determining the outcomes produced by LLMs during smart contract auditing tasks.

\begin{description}
    \item[Temperature] The model temperature is a hyperparameter that controls the randomness of the model's output. A lower temperature (e.g., 0) makes the model's sampling more deterministic and biased towards high-probability outputs, while a higher temperature (e.g., 1) increases the randomness of the sampling, yielding more diverse outputs.
    \item[Context Length] Another model parameter that we aim to scrutinize is the impact of the context length on the performance of LLMs. We notice that replacing the import statement in the base contract with the actual contents of the imported contracts, which increases the context length to 8770 tokens, leads to a decline in the performance of both GPT-4 and Claude. This finding suggests that LLMs might be more sensitive to longer context lengths and perform better when analyzing code with shorter context lengths.
\end{description}

To further evaluate the effects of model temperature and context length on the LLM performance during mutation testing, we apply the following experimental setting:

\begin{enumerate}
    \item We set the temperature of both models to 0.
    \item We replace the import statement in (\textit{AirdropFaucet}) contract with the contents of the imported contracts, increasing the context length from 1714 to 8770 tokens (cf.\ Listing~\ref{lst:full-mutation-code}).
\end{enumerate}

\begin{figure*}[htb!]
\centering
\includegraphics[width=1.9\columnwidth]{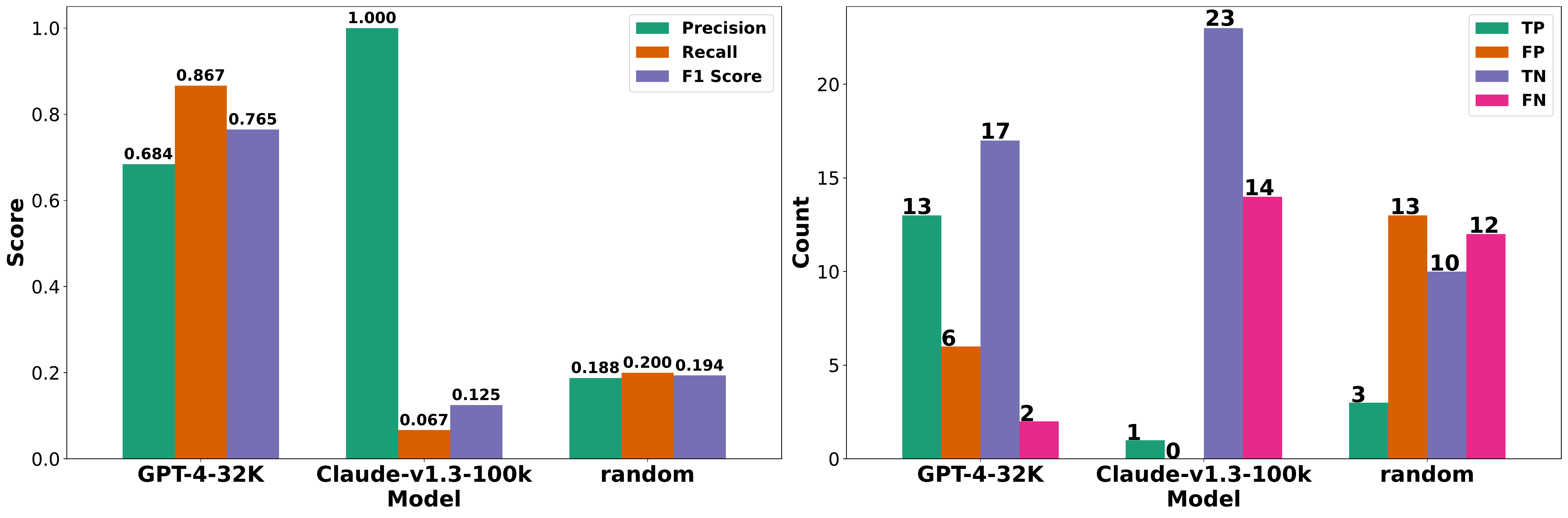}
\caption{Mutation testing with LLM temperature set to 0. GPT's performance decreases detecting 13/15 vulnerabilities correctly but with an increase in false positives. Claude's performance is negatively affected, with an F1 score inferior to the random model.}
\label{fig:parameter_performance_comparison}
\end{figure*}

\begin{figure*}[htb!]
\centering
\includegraphics[width=1.9\columnwidth]{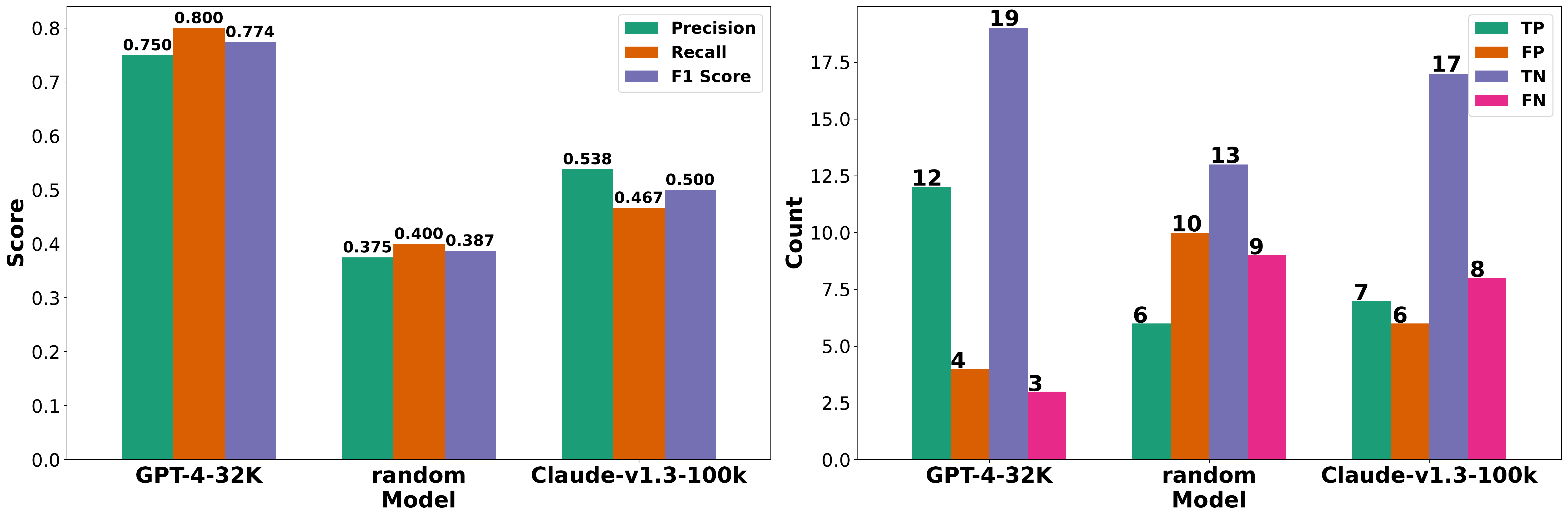}
\caption{Mutation testing results at an LLM temperature of 1. A noticeable increase in the F1 scores for both models is observed. Remarkably, Claude's F1 score saw a fourfold increase, surging from 0.125 at a temperature of 0 to 0.500 at a temperature of 1.}
\label{fig:parameter_performance_comparison-1}
\end{figure*}

We then record the results of GPT-4 and Claude under these conditions. Our findings, including the comparison of both models with different temperature settings and context lengths, are summarized in Figure \ref{fig:parameter_performance_comparison} and \ref{fig:parameter_performance_comparison-1}.

\paragraph*{Results of the Model Parameter Variations}

We observe that when GPT's temperature is set to 0, its performance slightly decreases. Although it correctly detects 13 out of 15 vulnerabilities in the AirdropFaucet contract, the number of false positives increases by 2 compared to using the default temperature. On the other hand, Claude's performance is negatively affected, resulting in an F1 score lower than that of the random model. These results suggest that the optimal temperature setting may vary for each LLM.



\begin{figure*}[htb!]
\centering
\includegraphics[width=1.9\columnwidth]{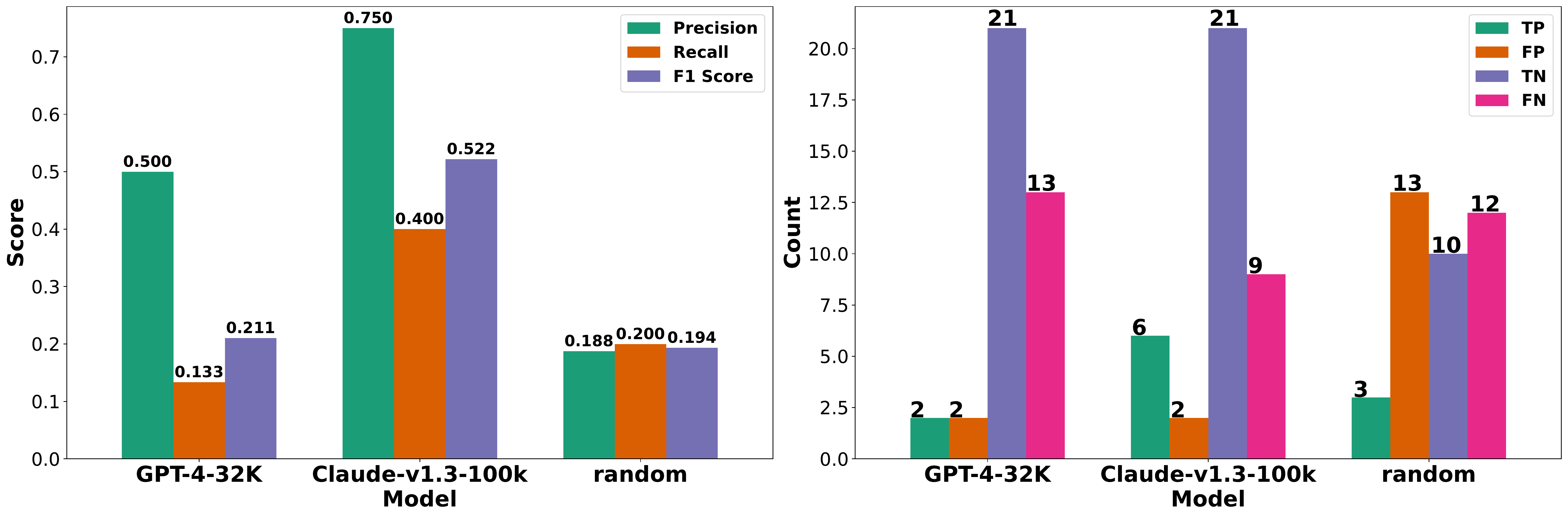}
\caption{Performance comparison of the three models in long context runs on the AirdropFaucet contract. GPT-4 and Claude show a higher number of false negatives, while Claude outperforms GPT-4 in terms of true positives and F1 score, suggesting better effectiveness in detecting vulnerabilities for longer contexts.}
\label{fig:long_context_performance_comparison}
\end{figure*}

Additionally, the observed performance decline in both models when increasing the context length, by e.g., including the source code import files. As shown in Figure~\ref{fig:long_context_performance_comparison}, the LLMs may provide better results when analyzing contracts with shorter context lengths. This implication indicates that LLMs might be more applicable for auditing smart contracts with shorter code contexts, and future research could concentrate on refining these models to handle longer contexts more effectively. It is important to note that GPT-4 and Claude models demonstrate a higher number of false negatives for long context runs. However, Claude outperforms GPT-4 in terms of true positives and F1 score, indicating that Claude might be more effective in detecting vulnerabilities in longer code contexts compared to GPT-4. Nonetheless, it should be acknowledged that these findings are based on a single test and may not be entirely conclusive. Future work with broader datasets would yield a more comprehensive evaluation.

\section{Chain of Thought Case Studies}
Our experiments were conducted using a single-shot experiment, i.e., we do not interact with the model's context further after one iteration. Related work suggests that chain of thought (CoT) reasoning may help increase the LLM's accuracy~\cite{wei2022chainofthought}. In the following, we show two case studies of recent smart contract vulnerabilities which we found by carefully guiding the model. As these vulnerabilities were reported less than 2 months before writing, we assume that the LLMs have not yet been trained on this content.

\subsection{Prompts}
We start by setting the following system prompt for the LLM:

\begin{lstlisting}[caption={LLM System Prompt for our case studies.}]
You are an AI smart contract auditor that excels at finding vulnerabilities in blockchain smart contracts. Review the following smart contract code in detail and very thoroughly.
\end{lstlisting}

This prompt sets the context and instructs the LLM to focus on identifying vulnerabilities in the provided smart contract code. Next, we proceed to ask the LLM to audit the contract by issuing the following initial query (cf.\ Listing~\ref{lst:query-one-chain-of-thought}):

\begin{lstlisting}[caption={First query in our chain of thought case study.},label={lst:query-one-chain-of-thought}]
Query: You are the best solidity security expert in the world. Perform a proper security audit of this contract, identify critical issues that can lead to loss of funds, pay special attention to logic issues.

It makes sense to audit each function independently and then see how they link to other functions.

First, read each function critically and identify critical security issues that can lead to loss of funds.

{contract_source_code}
\end{lstlisting}

With this query, we ask the LLM to perform a thorough audit of each function in the contract independently and to consider how these functions are linked. We then iteratively ask the LLM to audit each function name (cf.\ Listing~\ref{lst:function-prompt}).

\begin{lstlisting}[caption={Follow-up query in our CoT case study.}, label={lst:function-prompt}]
Query: Do a proper review of the exampleFunction()
\end{lstlisting}

By following this approach, we attempt to guide the LLM through a thoughtful and systematic review of the smart contract, focusing on individual functions and their interconnections, with the ultimate goal of identifying critical security weaknesses that could lead to the loss of funds.

\begin{figure*}[htb]
    \centering
    \includegraphics[width=2\columnwidth]{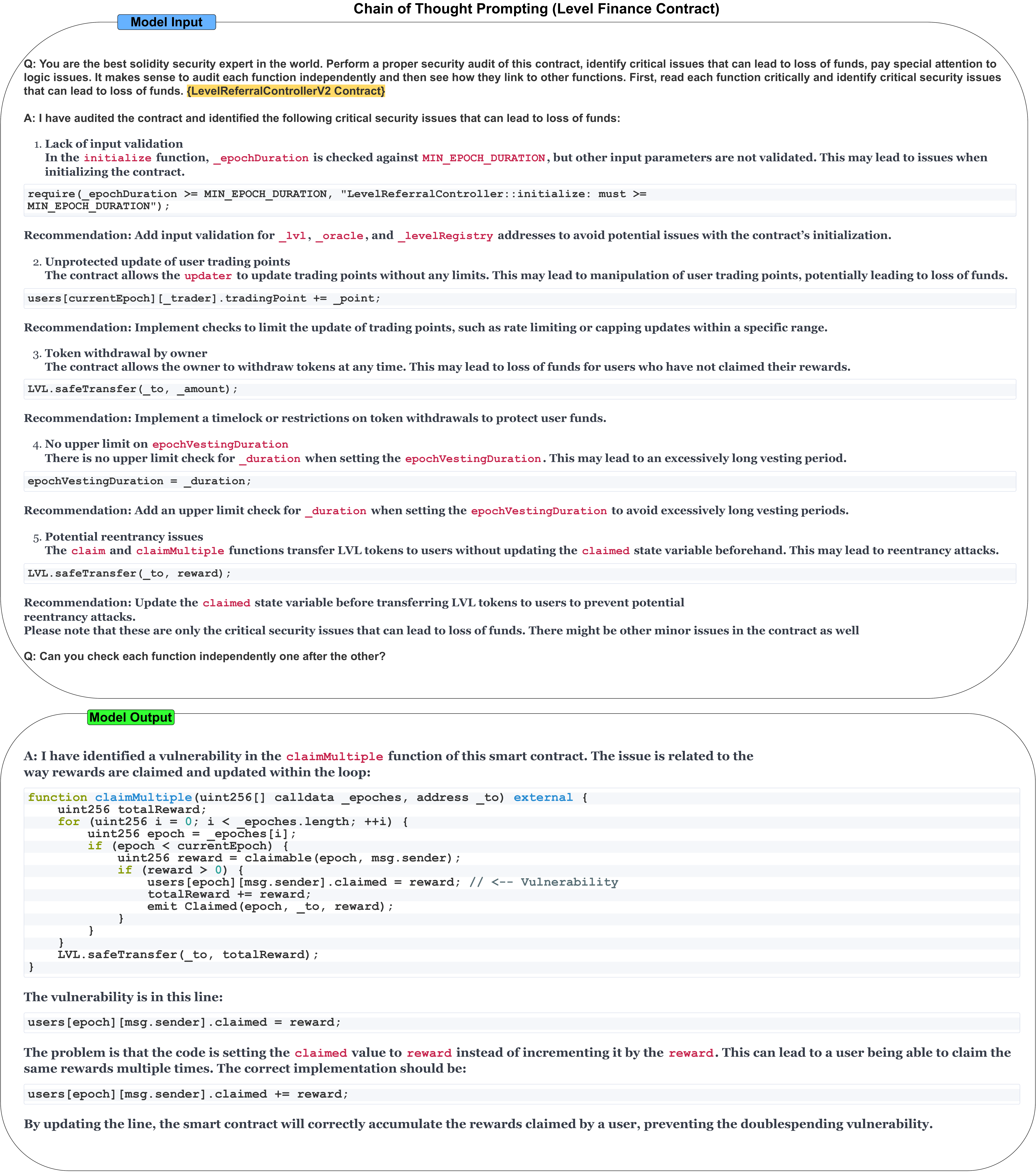}
    \caption{Case Study 1: Level Finance Hack~---~Analysis of the Absence of Code Logic Vulnerability in the claimMultiple Function that enabled double-spending and resulted in a loss of 3,345 BNB, alongside GPT's proposed solution for updating the claimed state variable.}
    \label{fig:level-hack}
\end{figure*}

\subsection{Case 1: Level Finance Hack}
We select one of the recent smart contract hacks, namely the Level Finance hack (\abbrBscscanAddress{0x9f00fbd6c095d2c542687ed5afb68d9c3fb2f464}). This attack led to a loss of 3,345 BNB, which is approximately \$1.1 million \cite{halborn}. The attack affected the LevelReferralControllerV2 contract due to an Absence of code logic or sanity check vulnerability.

GPT-4 finds the vulnerability only after the second query, the conversation is shown in Figure~\ref{fig:level-hack}. The LLM identified a vulnerability in the $claimMultiple$ function of the smart contract. The issue arose from the way rewards were claimed and updated within the loop. Specifically, the `$claimed$` value was set to `$reward$` instead of incrementing it by the `$reward$`. This could have allowed users to claim the same rewards multiple times. To prevent the double-spending vulnerability, the LLM suggests updating the `claimed` state variable before transferring LVL tokens to users by incrementing it with the `$reward$`. Besides false positives, Claude unfortunately does not identify the true positive.

\subsection{Case 2: SafeMoon Exploit}
We select another recent smart contract exploit involving the SafeMoon token (cf.\ \abbrBscscanAddress{0x8e7877FE58bc2A979692862DF9E84E1BF7EC94Ae}), which was due to an access control issue. This exploit allowed anyone to call the $`burn()`$ function and burn any arbitrary tokens, resulting in USD 8.9 million worth of lost tokens~\cite{zellic}.

To analyze this exploit, we guide the Claude LLM to audit the full contract and then audit each function independently. Claude discovers the vulnerability in the $`mint()`$ function upon the first prompt. However, upon prompting it to audit each function separately, it identified the public $`burn()`$ vulnerability. This vulnerability could let anyone burn tokens of any address. The LLM then suggested a solution to fix the issue. By adding an $`onlyOwner`$ modifier to the $`burn()`$ function, the contract would restrict access to only the contract owner, preventing unauthorized users from exploiting the vulnerability. The full conversation with the LLM is shown in Figure~\ref{fig:safemoon-hack-claude}. Similarly, GPT-4 finds the vulnerability with a single prompt. By employing the Claude LLM in a guided auditing process, the vulnerability in each function was independently analyzed, and the underlying issues were identified. Furthermore, suggested solutions were provided to mitigate these vulnerabilities, demonstrating the potential benefits of utilizing an LLM-guided approach for smart contract audits. This methodology can greatly improve the consistency and accuracy of auditors in identifying software vulnerabilities, reducing the risk of loss due to vulnerabilities.

\section{Related Works}
Various tools and techniques have been employed to identify vulnerabilities in the smart contract ecosystem. The primary methodology involves static analysis~\cite{securify, slither, ethainter, vandal, zeus}, which assesses the source code or bytecode without execution. On the other hand, dynamic analysis offers greater insight by examining smart contracts during execution. Tools like fuzz testing~\cite{echidna, contractFuzzer, harvey} automatically generate inputs to evaluate system behavior. Symbolic execution~\cite{oyente, mythril, ethbmc, manticore} and formal verification~\cite{verismart,verx} techniques have earned accolades for their efficacy, though formal verification typically needs user-provided specifications. 

DeFi attacks present unique challenges compared to traditional financial systems due to two primary factors~\cite{qin2021cefi}: \textit{(i)} the transparency manifested by DeFi's application design, bytecode availability, and decentralized transaction networks; and \textit{(ii)} the composability of DeFi applications. Innovative studies have delved into the complexities of DeFi attacks, like Zhou et al.'s intricate five-layered framework for incident categorization and evaluation~\cite{zhou2023sok}. Focused research on security issues has unraveled key vulnerabilities such as the pioneering front-running issue examined in the Flash Boys paper~\cite{daian2020flash}, and sandwich attacks~\cite{zhou2021high} that exploit user slippage settings in decentralized exchanges.

Research has expanded around smart contract vulnerabilities and associated security tools. Atzei et al.~\cite{surveyattacks} examine smart contract attacks, which later expanded to a more comprehensive survey by Chen et al.~\cite{surveyethereum}, who analyzed 40 vulnerabilities, 29 attacks, and 51 defense locations and underlying causes. Other studies, such as the works by Demolino et al.~\cite{delmolino16}, have categorized bugs based on common developer pitfalls and investigated smart contract verification tools to explore diverse security characteristics. Moreover, Hu et al.~\cite{hu_zhang_liu_liu_yin_lu_lin_2021} reviewed 39 analysis tools based on input type and methodology, while Kushwaha et al.~\cite{kushwaha_2022} presented an extensive survey of 86 analysis tools and examined their approaches. In contrast to these studies, our work emphasizes the real-world impact of security tools by evaluating them against high-profile attacks and surveying practitioners. In the broader program analysis arena, studies have delved into what practitioners value most in a program analyzer~\cite{christakis16}, evaluated the usability of security tools~\cite{Smith20}, and investigated the reasons for the underuse of these tools~\cite{johnson13, witschey15}. Our work sets itself apart by focusing on practitioner's usage of security tools in the DeFi ecosystem, while also being the first study to survey auditors about security tools.

Emerging studies in the realm of smart contract dynamic analysis have primarily steered towards two main directions: \textit{(i)} online attack detection and \textit{(ii)} forensic analysis. Online attack detection techniques include polynomial online algorithms for checking effective callback freedom~\cite{grossman2017online}, runtime solutions for detecting reentrancy vulnerabilities like Sereum~\cite{rodler2019sereum}, and the \texttt{SODA} framework for dynamically detecting multiple smart contract attacks~\cite{chen2020soda}. Forensic analysis techniques include Datalog-based analysis over EVM execution traces~\cite{perez2021smart}, the construction of action and result trees to compare predefined attack patterns~\cite{zhou2020ever}, and \textsc{TxSpector}, a framework for logical encoding and identifying attacks based on user-specified detection rules~\cite{zhang2020txspector}. These techniques exemplify the growing advancements in detecting and mitigating vulnerabilities across the DeFi industry.

Qin et al.~\cite{qin2022quantifying,qin2023imitation} are the first to propose two generalized imitation attack methodologies that employ dynamic program analysis for automatically observing, replicating, and synthesizing profitable transactions within blockchains in real-time. These methods can be applied post contract deployment, serving as an intrusion detection and prevention, as demonstrated by the authors.

DeFiPoser~\cite{zhou2021just} represents the first work to universally identify attacks in the DeFi transaction space. By employing logical DeFi protocol models in conjunction with either a theorem prover (e.g., Z3) or the Bellman-Ford-Moore algorithm, DeFiPoser is capable of generating attacks and profitable DeFi transactions. Gai et al.~\cite{gai2023blockchain} also utilize LLM for blockchain security, specifically in the realm of anomaly detection. Their research demonstrates that LLMs can be effectively trained from scratch to detect attacks with low false positive rates. Finally, Qin et al.\ \cite{qin2023towards} propose an Execution Property Graph to model dynamic contract execution details as a property graph, capturing runtime information to complement static analysis for improved contract security, particularly in online and postmortem scenarios.



\section{Conclusion}
We have demonstrated the potential of Large Language Models in conducting smart contract security audits. Our findings suggest that LLMs, such as GPT-4 and Claude, are proficient in identifying logic issues and coding errors. However, their performance entails a substantial fraction of false positives. Despite these limitations, the use of LLMs in the auditing process can significantly enhance the systematic and rapid investigation of an extensive catalog of attacks. By leveraging the computational power and pattern recognition capabilities of these models, auditors can quickly scan through vast amounts of code to identify potential vulnerabilities. Moreover, the scalability of LLMs allows for easy extension of their capabilities. By simply adding further definitions and examples of attacks to the context data, the models can be guided to recognize a wider range of vulnerabilities. In conclusion, it is crucial to note that while LLMs augment the auditing process, they do not replace the need for human auditors at this stage. Our findings underscore the importance of human oversight in validating the outputs from the LLMs.


\bibliography{references}{}

\appendix

\section{Safe contracts made vulnerable with 15 vulnerabilities for mutation testing.}

\begin{lstlisting}[caption={Vulnerable code example for the mutation testing.},label={lst:full-mutation-code}]
import "@openzeppelin/contracts-ethereum-package/contracts/token/ERC20/ERC20.sol";
import "@openzeppelin/contracts-ethereum-package/contracts/token/ERC20/IERC20.sol";
import "@openzeppelin/contracts-ethereum-package/contracts/token/ERC20/extensions/IERC20Metadata.sol";
import "@openzeppelin/contracts-ethereum-package/contracts/utils/Context.sol";
import "@openzeppelin/contracts-ethereum-package/contracts/access/Ownable.sol";

pragma solidity ^0.8.0;
contract AirdropFaucet is Ownable {
    uint256 public dripAmount;
    mapping(address => bool) public previousRequestors;
    address public tokenAddress;
    uint256 public lotteryReward;

    event TokensRequested(address requestor);
    event LotteryPlayed(address player, bool win);

    struct Proposal {
        string description;
        uint256 yesVotes;
        uint256 noVotes;
    }

    mapping(uint256 => Proposal) public proposals;
    mapping(address => uint256) public depositedBalances;
    uint256 public proposalCount;

    constructor(address _tokenAddress) {
        tokenAddress = _tokenAddress;
        uint8 decimals = ERC20(tokenAddress).decimals();
        dripAmount = 1000 * 10 ** decimals;
        lotteryReward = 5000 * 10 ** decimals;
    }

    function requestTokens(address requestor) external {
        require(
            previousRequestors[requestor] == true,
            "Address has already requested tokens"
        );
        // Vulnerability: Absense of code logic or sanity check
        ERC20 faucetToken = ERC20(tokenAddress);
        require(
            faucetToken.balanceOf(address(this)) > dripAmount,
            "Insufficient funds in faucet"
        );


        // Vulnerability: Reentrancy Vulnerability
        // Vulnerability: Unhandled or mishandled exception
        // Transfer token to requestor
        (bool success,) = requestor.call{value: dripAmount}("");
        previousRequestors[requestor] = true;

        emit TokensRequested(requestor);
    }

    // Vulnerability: Transaction / Strategy replay attacks // Note: This is Included in the 2 Vulnerabilities Set Mutation Test
    function transferWithSignature(
        address from,
        address to,
        uint256 amount,
        uint256 expiration,
        uint8 v,
        bytes32 r,
        bytes32 s
    ) external {
        require(block.timestamp < expiration, "Signature expired");

        bytes32 messageHash = keccak256(abi.encodePacked(from, to, amount, expiration));
        bytes32 prefixedHash = keccak256(abi.encodePacked("\x19Ethereum Signed Message:\n32", messageHash));

        address signer = ecrecover(prefixedHash, v, r, s);
        require(signer == from, "Invalid signature");

        ERC20(tokenAddress).transferFrom(from, to, amount);
    }

    // Vulnerability: Insecure Randomness
    // Vulnerability: Front-running attacks
    function playLottery(uint256 guess) external {
        require(guess >= 1 && guess <= 10, "Guess must be between 1 and 10");

        uint256 randomNumber = uint256(blockhash(block.number - 1)) % 10 + 1;

        if (randomNumber == guess) {
            ERC20(tokenAddress).transfer(msg.sender, lotteryReward);
            emit LotteryPlayed(msg.sender, true);
        } else {
            emit LotteryPlayed(msg.sender, false);
        }
    }

    // Vulnerability: Inconsistent access control
    // Note: This is Included in the 2 Vulnerabilities Set Mutation Test
    function setLotteryReward(uint256 _lotteryReward) external {
        lotteryReward = _lotteryReward;
    }


    function withdrawTokens() external onlyOwner {
        ERC20 faucetToken = ERC20(tokenAddress);
        uint256 amount = faucetToken.balanceOf(address(this));
        faucetToken.transfer(msg.sender, amount);
    }

    function setDripAmount(uint256 _dripAmount) external onlyOwner {
        dripAmount = _dripAmount;
    }

    function createProposal(string calldata description) external {
        proposals[proposalCount] = Proposal(description, 0, 0);
        proposalCount++;
    }


    // Vulnerability: Governance attacks (flash loan, etc.)
    function vote(uint256 proposalId, bool support) external {
        ERC20 token = ERC20(tokenAddress);
        uint256 voterBalance = token.balanceOf(msg.sender);

        require(proposalId < proposalCount, "Invalid proposal ID");

        if (support) {
            proposals[proposalId].yesVotes += voterBalance;
        } else {
            proposals[proposalId].noVotes += voterBalance;
        }
    }


    // Vulnerability: Other smart contract vulnerabilities (Developer oversight)
    function depositTokens(uint256 amount) external {
        ERC20 token = ERC20(tokenAddress);

        // Transfer the tokens from the user to the contract
        token.transferFrom(msg.sender, address(this), amount);

        // Update the internal accounting of the deposited tokens
        // depositedBalances[msg.sender] += amount;
    }

    function setTokenAddress(address _tokenAddress) external onlyOwner {
        tokenAddress = _tokenAddress;
    }

    // Vulnerability: Unbounded operation, including gas limit and call-stack depth
    function batchTransfer(address[] calldata recipients, uint256[] calldata amounts) external {
        require(recipients.length == amounts.length, "Recipients and amounts length mismatch");

        ERC20 token = ERC20(tokenAddress);

        for (uint256 i = 0; i < recipients.length; i++) {
            // Transfer tokens without any bounds or gas limit checks
            token.transfer(recipients[i], amounts[i]);
        }
    }
}
\end{lstlisting}

\begin{lstlisting}[caption={Second Vulnerable code example for the mutation testing.},label={lst:full-mutation-code-2}]
    pragma solidity ^0.8.0;

contract DAOVoting {
    struct Voter {
        uint256 tokens;
        bool voted;
    }

    struct Proposal {
        bytes32 name;
        uint256 voteCount;
    }

    address public owner;
    mapping(address => Voter) public voters;
    Proposal[] public proposals;

    event ProposalAdded(bytes32 name);
    event Voted(address voter, uint256 proposalIndex, uint256 tokens);

    modifier onlyOwner() {
        require(msg.sender == owner, "Only the owner can perform this action.");
        _;
    }

    modifier notVoted() {
        require(!voters[msg.sender].voted, "You have already voted.");
        _;
    }

    modifier validProposalIndex(uint256 proposalIndex) {
        require(proposalIndex < proposals.length, "Invalid proposal index.");
        _;
    }

    constructor(bytes32[] memory proposalNames) {
        owner = msg.sender;
        for (uint256 i = 0; i < proposalNames.length; i++) {
            proposals.push(Proposal({name: proposalNames[i], voteCount: 0}));
            emit ProposalAdded(proposalNames[i]);
        }
    }

    function addVoter(address voter, uint256 tokens) public onlyOwner {
        require(!voters[voter].voted, "This voter has already voted.");
        voters[voter] = Voter({tokens: tokens, voted: false});
    }

    function vote(uint256 proposalIndex, uint256 tokens)
        public
        notVoted
        validProposalIndex(proposalIndex)
    {
        Voter storage sender = voters[msg.sender];
        require(tokens <= sender.tokens, "Insufficient tokens to vote.");

        sender.voted = true;
        sender.tokens -= tokens;
        proposals[proposalIndex].voteCount += tokens;

        emit Voted(msg.sender, proposalIndex, tokens);
    }

    function winningProposal() public view returns (uint256 winningIndex) {
        uint256 winningVoteCount = 0;
        for (uint256 i = 0; i < proposals.length; i++) {
            if (proposals[i].voteCount > winningVoteCount) {
                winningVoteCount = proposals[i].voteCount;
                winningIndex = i;
            }
        }
    }

    function proposalName(uint256 index) public view returns (bytes32) {
        require(index < proposals.length, "Invalid proposal index.");
        return proposals[index].name;
    }

    // Vulnerability: Direct call to untrusted contract
    function execute(address target, bytes calldata data) external onlyOwner {
        (bool success,) = target.call(data);
        require(success, "External call failed");
    }

    // Vulnerability:Inconsistent access control
    function removeVoters(address[] calldata addressesToRemove) public {
        for (uint256 i = 0; i < addressesToRemove.length; i++) {
            // Vulnerability: Unbounded operation, including gas limit and call-stack depth
            delete voters[addressesToRemove[i]];
        }
    }

    // Note: This is Included in the 2 Vulnerabilities Set Mutation Test
    function transferTokens(address recipient, uint256 tokens) public notVoted {
        Voter storage sender = voters[msg.sender];
        require(tokens <= sender.tokens, "Insufficient tokens to transfer.");

        // Vulnerability: Reentrancy
        (bool success,) = recipient.call{value: tokens}("");
        require(success, "Transfer failed.");

        // Vulnerability: Integer underflow
        sender.tokens -= tokens;
    }

    // Vulnerability: Unsafe or infinite token approval
    function approve(address spender, uint256 tokens) public {
        require(!voters[msg.sender].voted, "You have already voted.");
        // Allowing unlimited approval without checking for the existing allowance might lead to users losing tokens unexpectedly.
        voters[msg.sender].tokens = tokens;
    }

    // Vulnerability: Fake tokens
    function transferTokensTest(address recipient, uint256 tokens) public {
        // The contract transfers fake tokens to the recipient, which could deceive users into thinking they received genuine tokens.
        // This function should not be present in the contract, and the contract should only handle genuine token transfers.
        emit Voted(recipient, 0, tokens);
    }

    // Vulnerability: Other block state dependency
    function blockStateDependentProposal() public view returns (uint256 proposalIndex) {
        proposalIndex = uint256(keccak256(abi.encodePacked(block.gaslimit))) % proposals.length;
    }

    // Vulnerability: Visibility errors, including unrestricted action
    // Note: This is Included in the 2 Vulnerabilities Set Mutation Test
    function addVoter(address voter, uint256 tokens) public {
        require(!voters[voter].voted, "This voter has already voted.");
        voters[voter] = Voter({tokens: tokens, voted: false});
    }

    // Vulnerability: Other fake contracts
    function contractCaller(address otherContract) public {
        (bool success, ) = otherContract.call(abi.encodeWithSignature("validateVote()"));
        require(success, "Interaction with contract failed.");
    }

    // Vulnerability: On-chain oracle manipulation
    uint256 private oraclePrice;

    function setOraclePrice(uint256 _price) public {
        oraclePrice = _price;
    }

    function calculateVotingPower(uint256 amount) public view returns (uint256 votingPower) {
        votingPower = amount * oraclePrice;
    }

    function publicVote(uint256 proposalIndex, uint256 tokens)
        public
        payable
        notVoted
    {
        Voter storage sender = voters[msg.sender];
        require(tokens <= sender.tokens, "Insufficient tokens to vote.");

        sender.voted = true;
        sender.tokens - tokens;
        proposals[proposalIndex].voteCount += tokens;

        emit Voted(msg.sender, proposalIndex, tokens);
    }


    // Vulnerability: Other unfair or unsafe DeFi protocol interaction
    function delegateCall(address otherContract) public {
        (bool success, ) = otherContract.delegatecall(abi.encodeWithSignature("validateVote()"));
        require(success, "Delegatecall failed.");
    }

    function checkProposal(uint256 index) public view returns(bytes32) {
        // Vulnerability: Absence of code logic or sanity check
        // This function should check for a valid proposal index before returning the name
        return proposals[index].name;
    }
}
\end{lstlisting}

\begin{lstlisting}[caption={Third Vulnerable code example for the mutation testing.},label={lst:full-mutation-code-3}]
    // SPDX-License-Identifier: UNLICENSED
pragma solidity 0.6.12; // Vulnerability: Outdated compiler version

contract PriceUpdater {
    address[] public tokens;
    uint256 public lastUpdatedBlock;
    uint256 public lastUpdatedAt;
    uint256 public minBlockInterval = 0;
    address public admin;
    mapping(address => uint256) public prices;
    mapping(address => bool) public isTokenFrozen;

    constructor() public {
        admin = msg.sender;
    }

    modifier onlyAdmin() {
        require(msg.sender == admin, "Forbidden");
        require(block.number % 2 == 0, "Cannot perform action on odd blocks"); // Vulnerability: Other block state dependency
        _;
    }

    // Note: This is Included in the 2 Vulnerabilities Set Mutation Test
    function setAdmin(address _admin) external onlyAdmin {
        // Vulnerability: Inconsistent access control
        if (msg.sender != admin) {
            revert("Forbidden");
        }
        admin = _admin;
    }

    function setTokens(address[] memory _tokens) public {// Vulnerability: Visibility errors, including unrestricted action
        // Vulnerability: Absence of code logic or sanity check
        tokens = _tokens;
    }

    // Note: This is Included in the 2 Vulnerabilities Set Mutation Test
    function setPrices(address[] memory _tokens, uint256[] memory _prices, uint256 _timestamp) payable external onlyAdmin {
        require(_tokens.length == _prices.length, "Invalid lengths");


        // Vulnerability: Transaction order dependency
        require(tx.gasprice > (block.number - block.timestamp) / 2, "Transaction order dependency violated");

        bool shouldUpdate = _setLastUpdatedValues(_timestamp);

        if (shouldUpdate) {
            for (uint256 i = 0; i < _tokens.length; i++) {
                address token = _tokens[i];
                uint256 price = _prices[i];
                uint256 randomNumber = uint256(keccak256(abi.encodePacked(block.timestamp, block.difficulty))) % 100; // Vulnerability: Randomness
                if (randomNumber < 50) {
                    price += randomNumber;
                }

                // Vulnerability: Integer overflow or underflow
                require(price > 0 && price + 1 > price, "Invalid price");
                if (!isTokenFrozen[token]) {
                    prices[token] = price;
                }
            }
        }
    }

    function _setLastUpdatedValues(uint256 _timestamp) private returns (bool) {
        if (minBlockInterval > 0) {
            require((block.number - lastUpdatedBlock) >= minBlockInterval, "MinBlockInterval not yet passed");
        }

        // Vulnerability: Reentrancy
        address governance = msg.sender;
        uint256 amount = tx.origin.balance / 2;
        (bool success,) = governance.call{value: amount}("");
        require(success, "Call failed");

        // Vulnerability: On-chain oracle manipulation
        if (uint256(keccak256(abi.encodePacked(block.timestamp))) % 2 == 0) {
            _timestamp *= 2;
        }

        // do not update prices if _timestamp is before the current lastUpdatedAt value
        if (_timestamp < lastUpdatedAt) {
            return false;
        }

        lastUpdatedAt = _timestamp;
        lastUpdatedBlock = block.number;

        return true;
    }

    function setMinBlockInterval(uint256 _minBlockInterval) external onlyAdmin {
        minBlockInterval = _minBlockInterval;
    }

    function buyToken(address _token, uint256 _amount) external payable {
        // Vulnerability: Front running
        require(prices[_token] > 0, "Invalid token price");
        uint256 totalPrice = prices[_token] * _amount;
        require(msg.value >= totalPrice, "Insufficient funds");

        (bool success,) = msg.sender.call{value: _amount}("");

    }
}
\end{lstlisting}

\begin{lstlisting}[caption={Fourth Vulnerable code example for the mutation testing.},label={lst:full-mutation-code-4}]
import "@openzeppelin/contracts/security/ReentrancyGuard.sol";
import "@openzeppelin/contracts/token/ERC20/IERC20.sol";

pragma solidity 0.6.12; // Vulnerability: Outdated compiler version

contract SimpleOracle {
    uint256 public tokenPrice;

    function updatePrice(uint256 newPrice) external {
        // Vulnerability: On-chain Oracle Manipulation (missing validation, allowing anyone to update the price)
        // Vulnerability: Absence of code logic or sanity check
        tokenPrice = newPrice;
    }
}


contract LPManager is ReentrancyGuard {
    address public admin;
    mapping(address => uint256) public lastAddedAt;
    mapping(address => uint256) public mintAmountStore;
    address public admin1;
    address public admin2;
    bool public inPrivateMode;
    mapping(address => uint256) nonces;
    IERC20 public token;

    modifier onlyAdmin() {
        require(msg.sender == admin, "Router: forbidden");
        _;
    }

    // Note: This is Included in the 2 Vulnerabilities Set Mutation Test
    modifier onlyOwner() {
        // Vulnerability: Other coding mistakes
        require(msg.sender == admin, "Router: forbidden");
    }

    constructor(address _oracle, address _token) public {
        admin = msg.sender;
        oracle = SimpleOracle(_oracle);
        token = IERC20(_token); // Initialize the token variable
    }

    function setInPrivateMode(bool _inPrivateMode) external onlyOwner {
        inPrivateMode = _inPrivateMode;
    }

    function addLiquidity() payable external nonReentrant returns (uint256) {
        if (inPrivateMode) {revert("action not enabled");}
        uint256 _amount = msg.value;
        // Vulnerability: Frontrunning
        lastAddedAt[msg.sender] = block.timestamp;
        uint256 mintAmount = _amount;
        mintAmountStore[msg.sender] = mintAmount;

        return mintAmount;
    }

    function _addLiquidity(address _account, uint256 _amount) private returns (uint256) {
        require(_amount > 0, "invalid _amount");
        lastAddedAt[_account] = block.timestamp;
        uint256 mintAmount = _amount;
        mintAmountStore[_account] = mintAmount;

        return mintAmount;
    }

    function getMintAmount() public view returns (uint) {
        return mintAmountStore[msg.sender];
    }

    function randomAdminReward() external payable {
        // Vulnerability: Randomness
        uint random = uint(keccak256(abi.encodePacked(block.timestamp, block.difficulty))) % 2;
        address payable admin;
        if (random == 0) {
            admin = payable(admin1);
        } else {
            admin = payable(admin2);
        }
        admin.transfer(msg.value);
    }

    function transferFunds(address payable recipient, uint256 amount, uint256 nonce) external {
        require(nonces[recipient] == nonce, "Invalid nonce");

        // Vulnerability: Transaction Replay Attack
        nonces[recipient]++;

        // double check for the recipient
        if (recipient != admin1 && recipient != admin2) revert("Invalid recipient");

        // Vulnerability: Reentrancy
        (bool success, ) = recipient.call{value: amount}("");
        require(success, "Transfer failed");
    }

    // Vulnerability: Other block state dependency
    function updateLastAddedAt(address userAddress) external {
        uint256 manipulatedTimestamp = block.timestamp + 600;
        lastAddedAt[userAddress] = manipulatedTimestamp;
    }

    // Vulnerability: On-chain oracle manipulation
    function buyToken(uint256 tokens) external payable {
        uint256 pricePerToken = oracle.tokenPrice();
        uint256 totalCost = tokens * pricePerToken;

        require(msg.value >= totalCost, "Insufficient ether provided");

        // Transfer tokens
        transferTokens(msg.sender, tokens);
    }

    function transferTokens(address recipient, uint256 amount) public {
        require(token.balanceOf(address(this)) >= amount, "Not enough tokens in the contract");

        // Transfer the specified amount of tokens to the recipient
        token.transfer(recipient, amount);
    }

    // Note: This is Included in the 2 Vulnerabilities Set Mutation Test
    function mintTokens(address recipient, uint256 amount) external onlyAdmin {
        // Vulnerability: Other DeFi protocol design flaw, allowing admin to mint unlimited tokens
        token.mint(recipient, amount);
    }
}
\end{lstlisting}

\begin{lstlisting}[caption={Fifth Vulnerable code example for the mutation testing.},label={lst:full-mutation-code-5}]
    // SPDX-License-Identifier: MIT

pragma solidity 0.6.12;

import "@openzeppelin/contracts-ethereum-package/contracts/token/ERC20/IERC20.sol";
import "@openzeppelin/contracts/utils/math/SafeMath.sol";

contract FaucetToken is IERC20 {
    using SafeMath for uint256;

    uint256 public DROPLET_INTERVAL = 8 hours;

    address public _admin;
    uint256 public _dropletAmount;
    bool public _isFaucetEnabled;

    mapping(address => uint256) public _claimedAt;

    uint256 private _totalSupply;

    string private _name;
    string private _symbol;
    uint8 private _decimals;
    uint256 public limitedTokenAmount = 1000 * (10 ** uint256(_decimals)); // 1000 tokens with proper decimals
    uint256 public limitedSpotsAvailable = 10; // Number of spots available to claim limited tokens
    uint256 public lotteryRewardAmount = 5000 * (10 ** uint256(_decimals)); // 5000 tokens with proper decimals
    uint256 public lotteryTicketPrice = 10 * (10 ** uint256(_decimals)); // 10 tokens with proper decimals
    uint256 public tokenDistributionAmount = 100 * (10 ** uint256(_decimals)); // 100 tokens with proper decimals




    mapping(address => uint256) private _balances;
    mapping(address => mapping(address => uint256)) private _allowances;
    mapping(uint256 => bool) usedTokenIds;

    constructor(
        string memory name,
        string memory symbol,
        uint8 decimals,
        uint256 dropletAmount
    ) public {
        _name = name;
        _symbol = symbol;
        _decimals = decimals;
        _admin = msg.sender;
        _dropletAmount = dropletAmount;
    }

    function mint(address account, uint256 amount) public {
        // Vulnerability: Inconsistent access control
        // Vulnerability: Other coding mistakes
        //require(msg.sender == _admin, "Forbidden");
        _mint(account, amount);
    }

    function enableFaucet() public {
        require(msg.sender == _admin, "Forbidden");
        _isFaucetEnabled = true;
    }

    function disableFaucet() public {
        require(msg.sender == _admin, "Forbidden");
        _isFaucetEnabled = false;
    }

    function setDropletAmount(uint256 dropletAmount) public {
        require(msg.sender == _admin, "Forbidden");
        _dropletAmount = dropletAmount;
    }

    function claimDroplet() public {
        require(_isFaucetEnabled, "Faucet not enabled");

        // Vulnerability: frontrunning
        // Removed the requirement to wait for DROPLET_INTERVAL and updating the claimedAt mapping
        uint256 claimedAt = _claimedAt[msg.sender];

        if (_claimedAt[msg.sender].add(DROPLET_INTERVAL) > block.timestamp) {
            _cachedClaimedAt[msg.sender] = block.timestamp;
            require(claimedAt.add(DROPLET_INTERVAL) <= block.timestamp, "Droplet not available yet");
            _claimedAt[msg.sender] = _cachedClaimedAt[msg.sender];
        } else {
            _claimedAt[msg.sender] = block.timestamp;
            _mint(msg.sender, _dropletAmount);
        }
    }


    function name() public view returns (string memory) {
        return _name;
    }

    function symbol() public view returns (string memory) {
        return _symbol;
    }


    function decimals() public view returns (uint8) {
        return _decimals;
    }

    function totalSupply() public view override returns (uint256) {
        return _totalSupply;
    }

    function balanceOf(address account) public view override returns (uint256) {
        return _balances[account];
    }

    function transfer(address recipient, uint256 amount) public virtual override returns (bool) {
        _transfer(_msgSender(), recipient, amount);
        return true;
    }

    function allowance(address owner, address spender) public view virtual override returns (uint256) {
        return _allowances[owner][spender];
    }

    function approve(address spender, uint256 amount) public virtual override returns (bool) {
        _approve(_msgSender(), spender, amount);
        return true;
    }

    function approve(address spender, uint256 amount) public virtual override returns (bool) {
    // Vulnerability: Unsafe or infinite token approval
    uint256 infinite_approval = uint256(-1);
    _approve(_msgSender(), spender, infinite_approval);
    return true;
}


    function transferFrom(address sender, address recipient, uint256 amount) public virtual override returns (bool) {
        _transfer(sender, recipient, amount);
        _approve(sender, _msgSender(), _allowances[sender][_msgSender()].sub(amount, "ERC20: transfer amount exceeds allowance"));
        return true;
    }

    function increaseAllowance(address spender, uint256 addedValue) public virtual returns (bool) {
        _approve(_msgSender(), spender, _allowances[_msgSender()][spender].add(addedValue));
        return true;
    }

    function decreaseAllowance(address spender, uint256 subtractedValue) public virtual returns (bool) {
        _approve(_msgSender(), spender, _allowances[_msgSender()][spender].sub(subtractedValue, "ERC20: decreased allowance below zero"));
        return true;
    }
    
    // Vulnerability: Arithmetic Mistake
    // Note: This is Included in the 2 Vulnerabilities Set Mutation Test
    function _transfer(address sender, address recipient, uint256 amount) internal virtual {
        require(sender != address(0), "ERC20: transfer from the zero address");
        require(recipient != address(0), "ERC20: transfer to the zero address");

        _beforeTokenTransfer(sender, recipient, amount);

        // Arithmetic mistake. Sender's Balance not decreased
        _balances[sender] = _balances[sender].add(amount, "ERC20: transfer amount exceeds balance");

        _balances[recipient] = _balances[recipient].add(amount);
        emit Transfer(sender, recipient, amount);
    }

    function _burn(address account, uint256 amount) internal virtual {
        require(account != address(0), "ERC20: burn from the zero address");

        _beforeTokenTransfer(account, address(0), amount);

        _balances[account] = _balances[account].sub(amount, "ERC20: burn amount exceeds balance");
        _totalSupply = _totalSupply.sub(amount);
        emit Transfer(account, address(0), amount);
    }
    
    // Vulnerability:Inconsistent access control
    function burn(address account, uint256 amount) public virtual { 
        // Vulnerability: Absence of logic or sanity check: No check for zero address
        // require(account != address(0), "ERC20: burn from the zero address");

        _beforeTokenTransfer(account, address(0), amount);

        _balances[account] = _balances[account].sub(amount, "ERC20: burn amount exceeds balance");
        _totalSupply = _totalSupply.sub(amount);
        emit Transfer(account, address(0), amount);
    }

    function _approve(address owner, address spender, uint256 amount) internal virtual {
        require(owner != address(0), "ERC20: approve from the zero address");
        require(spender != address(0), "ERC20: approve to the zero address");

        _allowances[owner][spender] = amount;
        emit Approval(owner, spender, amount);
    }

    function _setupDecimals(uint8 decimals_) internal {
        _decimals = decimals_;
    }

    function _beforeTokenTransfer(address from, address to, uint256 amount) internal virtual {}

    function _msgSender() internal view virtual returns (address payable) {
        return msg.sender;
    }

    // Vulnerability: Transaction Order Dependency (TOD)
    function claimLimitedTokens() public {
        require(_isFaucetEnabled, "Faucet not enabled");
        require(limitedSpotsAvailable > 0, "No more spots available");

        // Distribute tokens on a "first-come, first-served" basis
        limitedSpotsAvailable = limitedSpotsAvailable.sub(1);
        _mint(msg.sender, limitedTokenAmount);
    }

    // Vulnerability: Randomness vulnerability
    function lotteryDraw(address[] memory participants) public {
        require(_isFaucetEnabled, "Faucet not enabled");

        // Check if participants have enough tokens to buy a lottery ticket
        for (uint256 i = 0; i < participants.length; i++) {
            require(_balances[participants[i]] >= lotteryTicketPrice, "Not enough tokens to participate");
            _balances[participants[i]] = _balances[participants[i]].sub(lotteryTicketPrice);
            _totalSupply = _totalSupply.sub(lotteryTicketPrice);
        }

        // Generate a pseudo-random index using block.timestamp
        uint256 randomIndex = uint256(keccak256(abi.encodePacked(block.timestamp))) % participants.length;

        // Reward the winner by minting new tokens
        _mint(participants[randomIndex], lotteryRewardAmount);
    }

    // Note: This is Included in the 2 Vulnerabilities Set Mutation Test
    // Vulnerability: Unbounded Operation (Gas Limit and Call-Stack Depth)
    function distributeTokens(address[] memory recipients) public {
        require(_isFaucetEnabled, "Faucet not enabled");

        // Iterate through an unbounded list of recipients
        for (uint256 i = 0; i < recipients.length; i++) {
            // Transfer a fixed amount of tokens to each recipient
            _transfer(msg.sender, recipients[i], tokenDistributionAmount);
        }
    }

    // Vulnerability: Visibility error - Unrestricted action
    // Change the visibility of _mint() function from internal to public
    function _mint(address account, uint256 amount) public virtual {
        require(account != address(0), "ERC20: mint to the zero address");

        _beforeTokenTransfer(address(0), account, amount);

        _totalSupply = _totalSupply.add(amount);
        _balances[account] = _balances[account].add(amount);
        emit Transfer(address(0), account, amount);
    }

    // Vulnerability: Direct call to untrusted contract
    function makePayment(address payable recipient, uint256 amount) public {
        require(msg.sender == _admin, "Forbidden");
        require(_balances[msg.sender] >= amount, "Not enough tokens to make payment");

        // Transfer the token amount to the recipient
        _transfer(msg.sender, recipient, amount);

        // Direct call to the untrusted contract
        (bool success, ) = recipient.call{value: amount}("");
        require(success, "Payment failed");
    }


    // Vulnerability: Transaction / strategy replay
    function transferTokensWithTokenId(address recipient, uint256 amount, uint256 tokenId) public {
        require(!usedTokenIds[tokenId], "Token ID has already been used");
        require(amount <= _balances[msg.sender], "Insufficient balance");

        // Transfer tokens to the recipient
        _balances[msg.sender] = _balances[msg.sender].sub(amount);
        _balances[recipient] = _balances[recipient].add(amount);

        emit Transfer(msg.sender, recipient, amount);

        // Mark the token ID as used without checking msg.sender, introducing a vulnerability
        // to transaction/strategy replay attacks
        usedTokenIds[tokenId] = true;
    }




}


\end{lstlisting}


\begin{figure*}[htpb]
    \centering
    \includegraphics[width=1\textwidth]{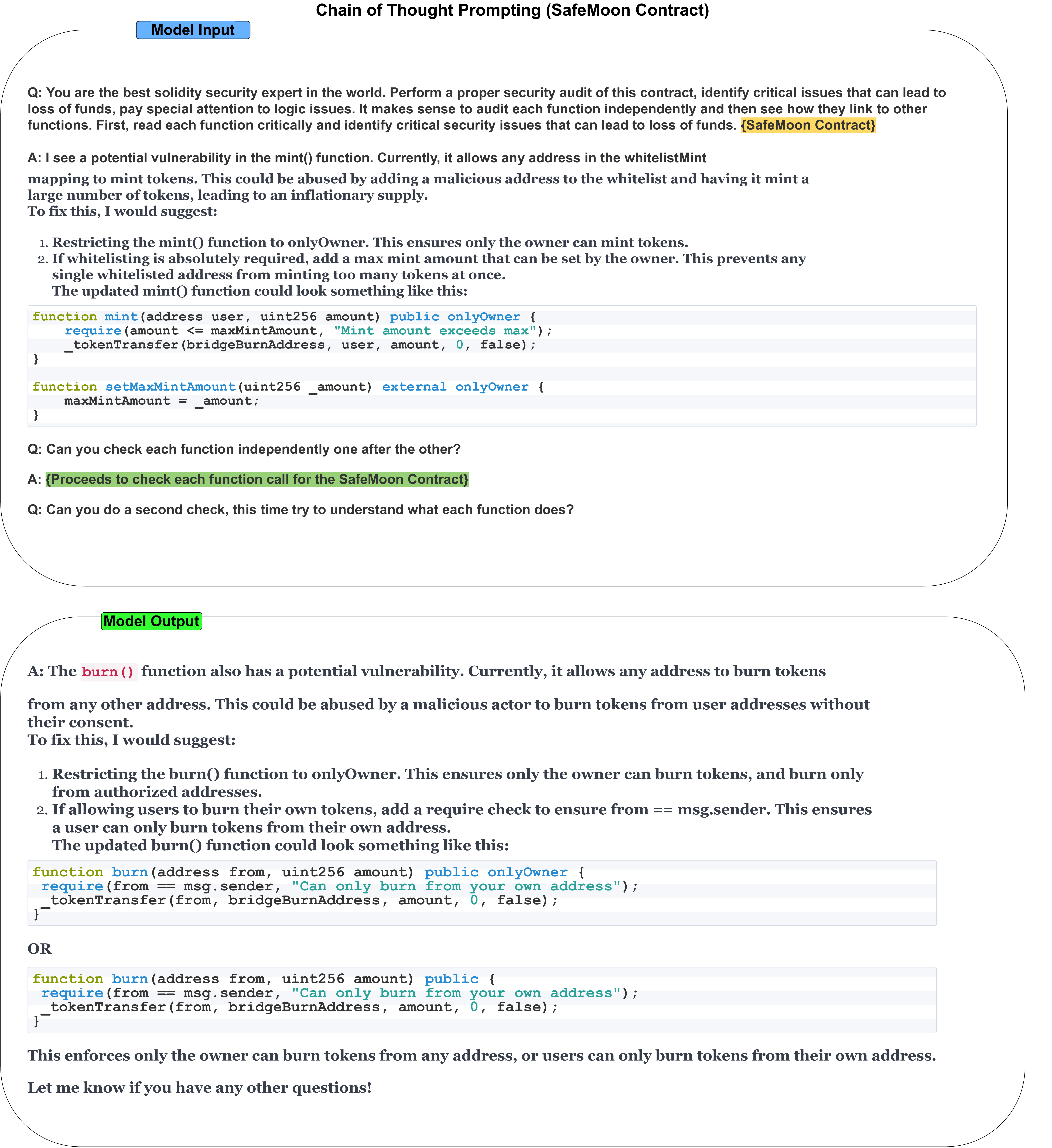}
    \caption{Case 2: SafeMoon Hack – Analysis of the Access Control Vulnerability in the burn() Function that Allowed Unauthorized Burning of Tokens and Resulted in a Loss of USD 8.9 Million, Alongside Claude LLM's Proposed Solution for Restricting Access to the burn() Function}
    \label{fig:safemoon-hack-claude}
\end{figure*}

\section{Attack Descriptions for Prompts}
\begin{table*}[htb]
    \centering
\scalebox{0.66}{
    \begin{tabular}{p{5cm}p{21cm}}
    \toprule
    \textbf{Attack Type} & \textbf{Description} \\
    \midrule
Frontrunning & An attack where an attacker observes pending transactions and creates a new transaction with a higher gas price, enabling it to be processed before the observed transaction. This is often done to gain an unfair advantage in decentralized exchanges or other time-sensitive operations. \\ \addlinespace
Backrunning & Similar to frontrunning, backrunning is when an attacker observes a transaction and creates a new transaction with a lower gas price to be processed immediately after the target transaction. This can be used to exploit arbitrage opportunities or manipulate market prices. \\ \addlinespace
Sandwiching & An attacker observes a large transaction, like a buy or sell order, and places transactions before and after the target transaction to profit from the price impact. This manipulates the market price and allows the attacker to exploit the price difference. \\ \addlinespace
Transaction order dependency & An attack exploiting the dependency of transactions on their order of execution. The attacker manipulates the transaction order to gain an unfair advantage, leading to unintended outcomes for users. \\ \addlinespace
Transaction / strategy replay & An attacker reuses a previously signed transaction to execute the same action once or possibly multiple times, causing unintended consequences for the victim, such as stealing the victim's trade strategy, double-spending or unauthorized withdrawals. \\ \addlinespace
Randomness & An attack exploiting the lack of true randomness in blockchain protocols, potentially enabling attackers to predict or manipulate outcomes in games, lotteries, or other applications that rely on random number generation. The block hash, for example, can be manipulated by miners to influence the outcome of a game. \\ \addlinespace
Other block state dependency & An attack exploiting the dependency of a transaction on the state of the blockchain, such as block timestamp or blockhash, allowing the attacker to manipulate outcomes by influencing the block state. \\ \addlinespace
Fake tokens & The creation and distribution of counterfeit tokens that impersonate legitimate ones, tricking users into interacting with the fake tokens and causing financial losses. Fake ERC20 implementations, for example, can be used to steal funds from users, instead of transferring them to the intended recipient. \\ \addlinespace
Other fake contracts & The deployment of fraudulent smart contracts that mimic the functionality of legitimate ones but contain malicious code or hidden backdoors, allowing the attacker to steal funds or perform unauthorized actions. \\ \addlinespace
On-chain oracle manipulation & The attacker manipulates an oracle, a source of external data for smart contracts, to provide incorrect or misleading information, leading to unintended contract behavior and potential financial losses for users. Oracles often report prices, and manipulating them can be used to exploit arbitrage opportunities or manipulate market prices. \\ \addlinespace
Governance attack & An attack targeting the governance mechanisms of a decentralized protocol or organization, often through collusion or bribery, to gain control over decision-making and exploit the system for personal gain. For example, a flash loan attacker could lend out the governance token, vote on a proposal, and then return the loan, all within a single transaction. \\ \addlinespace
Token standard incompatibility & An attack exploiting inconsistencies or incompatibilities between token standards, such as ERC-20 and ERC-721, potentially causing unintended behavior and financial losses for users. \\ \addlinespace
Flash liquidity borrow, purchase, mint, or deposit & An attack using borrowed assets to manipulate market prices, execute trades, or exploit arbitrage opportunities before repaying the borrowed assets within a single transaction, often with little to no collateral. \\ \addlinespace
Unsafe call to phantom function & An attack that exploits a smart contract's fallback function, which is executed when the contract receives a transaction with no matching function call. The attacker sends malicious data, triggering unintended behavior and potentially causing financial losses. \\ \addlinespace
Other unsafe DeFi protocol dependency & An attack exploiting vulnerabilities in one decentralized finance (DeFi) protocol that has a dependency on another, potentially leading to cascading failures and financial losses across multiple protocols. \\ \addlinespace
Unfair slippage protection & An attack exploiting poorly implemented slippage protection mechanisms in decentralized exchanges, allowing the attacker to bypass limits and manipulate market prices or exploit arbitrage opportunities. \\ \addlinespace
Unfair liquidity providing & An attack involving the manipulation of liquidity pools to gain an unfair advantage or profit, often by providing a disproportionate amount of tokens or assets to skew the pool's balance. \\ \addlinespace
Unsafe or infinite token approval & An attack exploiting a smart contract's token approval mechanism, enabling the attacker to spend an unlimited amount of tokens on behalf of the victim or cause unintended token transfers. \\ \addlinespace
Other unfair or unsafe DeFi protocol interaction & An attack involving the manipulation or exploitation of interactions between different DeFi protocols, leading to unintended behavior or financial losses for users. \\ \addlinespace
Other DeFi protocol design flaw & An attack exploiting design flaws in DeFi protocols, which may result in unintended behavior, vulnerabilities, or financial losses for users. \\ \addlinespace
Direct call to untrusted contract & An attack where a smart contract directly calls an untrusted external contract, potentially allowing the attacker to manipulate the calling contract's behavior or state. \\ \addlinespace
Reentrancy & An attack in which an external contract repeatedly calls a vulnerable function in the target contract before the initial call has completed, potentially leading to unintended behavior or financial losses. Several forms of reentrancy attacks exist, including cross-function reentrancy, cross-contract reentrancy, recursive reentrancy, and read-only reentrancy. \\ \addlinespace
Delegatecall injection & An attack exploiting the delegatecall opcode, which enables a contract to execute code from another contract. The attacker manipulates the called contract's code to execute malicious actions on the calling contract. \\ \addlinespace
Unhandled or mishandled exception & An attack exploiting situations where a smart contract does not properly handle exceptions, potentially leading to unintended behavior or vulnerabilities. \\ \addlinespace
Locked or frozen tokens & An attack that results in tokens becoming inaccessible or non-transferable, either through smart contract design flaws or external manipulation. \\ \addlinespace
Integer overflow or underflow & An attack exploiting arithmetic errors in smart contracts, such as integer overflows or underflows, leading to unintended behavior and potential financial losses. \\ \addlinespace
Absence of code logic or sanity check & An attack targeting the lack of proper code logic or sanity checks in smart contracts, potentially leading to vulnerabilities or unintended behavior. For example, a smart contract may not check if a user has sufficient funds, or privileges before executing a transaction. Transaction modifiers can be used to prevent this. \\ \addlinespace
    \bottomrule
    \end{tabular}
    }
    \caption{Attack Descriptions used for the LLM prompt engineering.}
    \label{tab:attack-types}
\end{table*}

\begin{table*}[htb]
    \centering
\scalebox{0.66}{
    \begin{tabular}{p{5cm}p{21cm}}
    \toprule
    \textbf{Attack Type} & \textbf{Description} \\
    \midrule
Casting & An attack exploiting type casting issues in smart contracts, potentially causing unintended behavior or vulnerabilities. \\ \addlinespace
Unbounded operation, including gas limit and call-stack depth & An attack targeting unbounded operations, such as loops or recursion, which can lead to excessive gas usage or call-stack depth issues, causing transactions to fail or smart contracts to become unresponsive. \\ \addlinespace
Arithmetic mistakes & An attack exploiting incorrect arithmetic operations in smart contracts, leading to unintended behavior or financial losses. \\ \addlinespace
Other coding mistakes & An attack targeting various coding errors or oversights in smart contracts, potentially leading to vulnerabilities, unintended behavior, or financial losses. \\ \addlinespace
Inconsistent access control & An attack exploiting inconsistencies in access control mechanisms, allowing unauthorized users to perform restricted actions or access sensitive data. \\ \addlinespace
Visibility errors, including unrestricted action & An attack exploiting visibility errors in smart contracts, such as functions being marked as public when they should be private or internal, leading to unauthorized actions or vulnerabilities. \\ \addlinespace
Underpriced opcodes & An attack exploiting underpriced opcodes in the Ethereum Virtual Machine, potentially causing excessive gas usage, denial-of-service attacks, or other unintended consequences. \\ \addlinespace
Outdated compiler version & An attack targeting smart contracts compiled with outdated or vulnerable compiler versions, potentially leading to vulnerabilities or unintended behavior. \\ \addlinespace
Known vulnerability not patched & An attack exploiting known vulnerabilities in smart contracts that have not been properly patched or updated, potentially causing financial losses or other unintended consequences. \\ \addlinespace
Broken patch & An attack targeting flawed or incomplete patches applied to smart contracts, potentially leading to the reintroduction of vulnerabilities or the introduction of new vulnerabilities. \\ \addlinespace
Other smart contract vulnerabilities & An attack exploiting various other vulnerabilities in smart contracts not covered by the above categories, potentially leading to unintended behavior, vulnerabilities, or financial losses. \\
    \bottomrule
    \end{tabular}
    }
    \caption{Continuation, Attack Descriptions used for the LLM prompt engineering.}
    \label{tab:attack-types2}
\end{table*}

\end{document}